\pdfoutput=1

\documentclass[11pt]{article}

\usepackage[final]{acl}

\usepackage{times}
\usepackage{latexsym}

\usepackage[T1]{fontenc}

\usepackage[utf8]{inputenc}

\usepackage{microtype}

\usepackage{inconsolata}

\usepackage{graphicx}
\usepackage{booktabs} 
\usepackage{array}    
\usepackage{multirow} 
\usepackage{threeparttable} 
\usepackage{caption}        
\usepackage[utf8]{inputenc} 
\usepackage{amsmath,amssymb,bm}
\usepackage{tabularx}
\usepackage{longtable}

\title{TF-Mamba: Text-enhanced Fusion Mamba with Missing Modalities for Robust Multimodal Sentiment Analysis}

\author{
  Xiang Li$^{1}$\thanks{Equal contribution.}, 
  Xianfu Cheng$^{2}$\footnotemark[1], 
  Dezhuang Miao$^{1}$, 
  Xiaoming Zhang$^{1}$\thanks{Corresponding author.}, 
  Zhoujun Li$^{2}$ \\
  $^{1}$School of Cyber Science and Technology, Beihang University, Beijing \\
  $^{2}$School of Computer Science and Engineering, Beihang University, Beijing \\
  \texttt{\{xlggg, buaacxf, taiyue, yolixs, lizj\}@buaa.edu.cn} \\
}

\begin{document}
\maketitle
\begin{abstract}
Multimodal Sentiment Analysis (MSA) with missing modalities has attracted increasing attention recently.
While current Transformer-based methods leverage dense text information to maintain model robustness, their quadratic complexity hinders efficient long-range modeling and multimodal fusion.
To this end, we propose a novel and efficient \textbf{T}ext-enhanced \textbf{F}usion \textbf{Mamba} (TF-Mamba) framework for robust MSA with missing modalities.
Specifically, a Text-aware Modality Enhancement (TME) module aligns and enriches non-text modalities, while reconstructing the missing text semantics.
Moreover, we develop Text-based Context Mamba (TC-Mamba) to capture intra-modal contextual dependencies under text collaboration.
Finally, Text-guided Query Mamba (TQ-Mamba) queries text-guided multimodal information and learns joint representations for sentiment prediction.
Extensive experiments on three MSA datasets demonstrate the effectiveness and efficiency of the proposed method under missing modality scenarios.
Code is available at \url{https://github.com/codemous/TF-Mamba}.
\end{abstract}

\section{Introduction}
Multimodal Sentiment Analysis (MSA) aims to understand and integrate sentiment cues expressed in multiple modalities (e.g., text, visual, and audio). 
Previous studies \citep{yang2022disentangled,sun2022cubemlp,li2023decoupled,feng2024knowledge} demonstrate that integrating complementary multimodal information gains MSA performance and enables its vital role in several applications.
However, most existing methods \citep{han2021improving,mai2022hybrid,zhang2023learning,wang2024cross,zeng2024disentanglement} are developed under ideal laboratory conditions where all modalities are assumed to be fully available during both training and inference.
In real-world scenarios, many inevitable factors like background noise and sensor failures often result in incomplete or corrupted modalities. 
These challenges severely undermine the performance and robustness of MSA models in practice. 

Recent studies \citep{yuan2021transformer,yuan2023noise,li2024unified,zhang2024towards} attempt to tackle the challenge of missing modalities in MSA. 
Among them, Transformer-based fusion models achieve notable progress owing to their powerful sequence modeling capabilities. 
For example, TFR-Net \citep{yuan2021transformer} adopts a Transformer-based feature reconstruction strategy to recover missing information in multimodal sequences.
More recently, LNLN \citep{zhang2024towards} prioritizes high-quality text features and treats text as the dominant modality to improve model robustness under various noise conditions. Unfortunately, these approaches often suffer from quadratic computational complexity and fail to achieve efficient text-enhanced multimodal fusion.
Mamba \citep{gu2023mamba}  emerges as a linear-time alternative to Transformers, exhibiting great promise in various domains \citep{yang2024cardiovascular,li2024coupled}.
Nevertheless, incorporating text enhancement into Mamba-based architectures for efficient missing modality modeling and fusion remains unexplored, which could potentially contribute to enhancing  MSA performance.

To address these issues, we propose a novel and efficient \textbf{T}ext-enhanced \textbf{F}usion \textbf{Mamba} (TF-Mamba) framework for robust MSA with missing modalities. 
TF-Mamba features three core text-dominant components: Text-aware Modality Enhancement (TME), Text-based Context Mamba (TC-Mamba), and Text-guided Query Mamba (TQ-Mamba).
Specifically, the TME module first aligns and enhances audio and visual modalities with text features, and reconstructs missing textual semantics from incomplete inputs. 
Subsequently, TC-Mamba leverages text information as the collaborative bridge within Mamba to model contextual dependencies in audio and video streams. The collaborative signals refine text representations in return for capturing shared semantics. 
Finally, TQ-Mamba queries informative multimodal features via text-guided cross-attention. 
It then learns cross-modal interactions with Mamba blocks to generate joint representations for sentiment prediction. The contributions are summarized as follows:
\begin{itemize}
    \item  To our knowledge, TF-Mamba is the first attempt incorporating text enhancement into Mamba-based fusion architecture for robust MSA under missing modality conditions. 
    
    \item We design three text-dominant modules to efficiently learn intra-modal dependencies and cross-modal interactions from incomplete modalities, thereby enhancing model robustness with dense text sentiment information.
    
    \item Extensive experiments on three MSA benchmarks demonstrate the superiority of TF-Mamba under uncertain missing modality scenarios.
    For instance, on MOSI, TF-Mamba outperforms Transformer-based state-of-the-art with fewer parameters and FLOPs.
\end{itemize}

\section{Related Work}
\subsection{Multimodal Sentiment Analysis}
Multimodal Sentiment Analysis (MSA) integrates multimodal information to understand and analyze human sentiments.
Mainstream studies in MSA \citep{li2024adaptive,li2025multi,li2025learning,sun2025sequential} focus on developing sophisticated fusion strategies and interaction mechanisms to improve sentiment prediction performance.
For example, \citet{han2021improving} hierarchically maximize mutual information between unimodal pairs to enhance multimodal fusion.
\citet{zhang2023learning} leverage sentiment-intensive cues to guide the representation learning of other modalities.
Despite these progresses, most methods assume fully available modalities, which is rarely achievable in real-world scenarios due to random data missing issues.

Recent studies \citep{yuan2021transformer,yuan2023noise,li2024unified,li2024toward,zhang2024towards,guo2024multimodal}  handle missing modalities by learning joint multimodal representations from available data or reconstructing incomplete information.
For instance, \citet{li2024unified} decompose modalities into sentiment-relevant and modality-specific components to reconstruct sentiment semantics through modality translation. 
However, they overlook the crucial role of text in sentiment expression without ensuring the quality of
dominant modality representations.
Recognizing this, \citet{zhang2024towards} propose a Transformer-based language-dominated noise-resistant network that prioritizes textual features to improve robustness under noisy conditions.
However, Transformer-based methods suffer from considerable computational overhead due to their inherent quadratic complexity, making them inefficient for modeling long or high-dimensional sequences. 
Given the rich sentiment cues in text and Mamba's efficiency in modeling long-range dependencies, our work explores text-enhanced strategies within the efficient Mamba architectures for robust MSA with random missing modalities.

\subsection{State Space Models and Mamba}
State Space Models (SSMs) \citep{gu2022parameterization,GuGR22} recently gain considerable attention for their ability to capture long-range dependencies with linear computational complexity.
Mamba~\citep{gu2023mamba} as an extension of SSMs, incorporates selection mechanisms and hardware-aware parallel algorithms to enhance sequence modeling efficiency. 
Its parallel scanning strategy further accelerates inference and achieves impressive performance on long-sequence tasks across various domains~\citep{ZhuL0W0W24, yang2024cardiovascular, jiang2025dual}.

Building on Mamba’s success, several studies ~\citep{li2024coupled,li2024spmamba, dong2025fusion,li2025alignmamba} explore its potential for multimodal fusion.
For instance, \citet{qiao2024vl} concatenate visual and textual sequences before jointly modeling them through Mamba blocks.
\citet{li2025alignmamba} learns both local and global cross-modal alignments prior to Mamba-based multimodal fusion.  
Recently, \citet{ye2025depmamba} pioneer the use of Mamba for depression detection, combining hierarchical modeling with progressive fusion strategies.  
Overall, these Mamba-based frameworks deliver competitive performance and improved efficiency compared to Transformer-based methods.
However, its application to MSA with missing modalities remains underexplored, especially in integrating text-enhanced fusion strategies into Mamba’s efficient modeling architecture. 
In contrast, we treat text as dominant modality during Mamba-based sequence modeling and multimodal fusion to enhance model robustness in incomplete modality settings.
\begin{figure*}[t]
  \includegraphics[width=\textwidth]{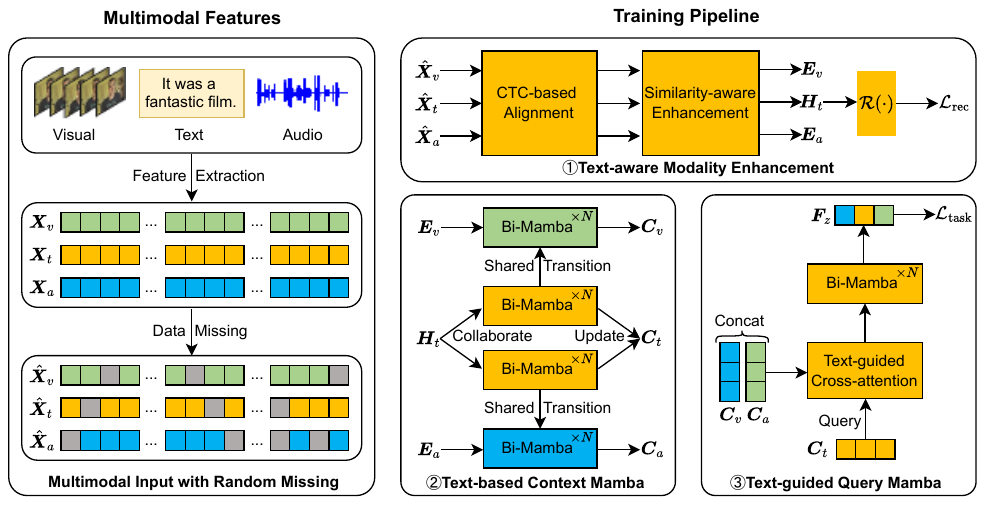}
  \caption{Overview of the TF-Mamba framework, which consists of three main components: Text-aware Modality Enhancement (TME), Text-based Context Mamba (TC-Mamba), and Text-guided Query Mamba (TQ-Mamba). Yellow blocks indicate the dominant role of the text modality in the training pipeline.}
  \label{fig:model}
\end{figure*}
\section{Method}
\subsection{Preliminary of Mamba}
State Space Models (SSMs) \cite{gu2022parameterization,GuGR22} gain increasing attention in recent years. 
An SSM maps a 1D sequence $\bm{x}(t) \in \mathbb{R}^{L}$ to $\bm{y}(t) \in \mathbb{R}^{L}$ via a hidden state $\bm{h}(t) \in \mathbb{R}^{N}$, where $N$ and $L$ denote the number of hidden states and sequence length. 
The above process is defined as:
\begin{align}
    \bm{h}'(t) &= \mathbf{A}\bm{h}(t) + \mathbf{B}\bm{x}(t) \label{eq:continuous_ssm} \\
    \bm{y}(t) &= \mathbf{C}\bm{h}(t) \label{eq:output}
\end{align}
where $\mathbf{A} \in \mathbb{R}^{N \times N}$ denotes the state transition matrix, and $\mathbf{B} \in \mathbb{R}^{N \times 1}$ and $\mathbf{C} \in \mathbb{R}^{1 \times N}$ are the input and output projections.

To make the continuous-time SSM system suitable for digital computing and real-world data, Mamba employs the zero-order hold method to discretize the continuous parameters $\mathbf{A}$ and $\mathbf{B}$ into $\overline{\mathbf{A}}$ and $\overline{\mathbf{B}}$ using a time scale parameter $\Delta$:
\begin{align}
    \overline{\mathbf{A}} &= \exp(\Delta \mathbf{A}) \label{eq:discrete_A} \\
    \overline{\mathbf{B}} &= (\Delta \mathbf{A})^{-1} \left( \exp(\Delta \mathbf{A}) - \mathbf{I} \right) \cdot \Delta \mathbf{B} \label{eq:discrete_B}
\end{align}
The resulting discretized system with step size $\Delta$ can then be formulated as:
\begin{align}
    \bm{h}_t &= \overline{\mathbf{A}} \bm{h}_{t-1} + \overline{\mathbf{B}} \bm{x}_t \label{eq:recurrent} \\
    \bm{y}_t &= \mathbf{C} \bm{h}_t \label{eq:discrete_output}
\end{align}
Mamba further unfolds the hidden states recursively and computes the output as:
\begin{align}
    \overline{\mathbf{K}} &= \left( \mathbf{C} \overline{\mathbf{B}}, \mathbf{C} \overline{\mathbf{A}} \overline{\mathbf{B}}, \ldots, \mathbf{C} \overline{\mathbf{A}}^{L-1} \overline{\mathbf{B}} \right) \label{eq:conv_kernel} \\
    \bm{y} &= \bm{x} \circledast \overline{\mathbf{K}} \label{eq:convolution}
\end{align}
where $\circledast$ denotes the convolution operation, and $\overline{\mathbf{K}} \in \mathbb{R}^{L}$ is the global convolution kernel. 
Mamba introduces the content-aware selection mechanism to modulate state transitions, enhancing its capacity to model complex temporal dependencies.

\subsection{Framework Overview}
The objective of our robust MSA task is to predict sentiment from video clips under scenarios where one or more modality features may be partially missing. As illustrated in Figure~\ref{fig:model}, the framework first generates multimodal features with randomly missing data, establishing the foundation for TF-Mamba training.
Following input preparation, the TME module performs text-centric CTC-based unimodal alignment, conducts similarity-aware token enhancement, and reconstructs missing textual semantics. The standardized multimodal features are then fed into the TC-Mamba and TQ-Mamba modules for efficient intra-modal modeling and cross-modal fusion,  generating robust joint multimodal representations for sentiment prediction.
The following sections provide a detailed description of each component in TF-Mamba.
\subsection{Multimodal Input with Random Missing}
Following prior work~\citep{yuan2021transformer,zhang2024towards}, we obtain the initial embeddings $ \bm{X}_m \in \mathbb{R}^{T_m \times D_m} $ using standard toolkits, where $ m \in \{t, v, a\} $ represents the text, visual, and audio modalities, respectively. Here, $ T_m $ denotes the sequence length and $ D_m $ is the feature dimension.

To simulate incomplete multimodal scenarios, we apply random replacement with missing values to the input sequences, resulting in corrupted inputs $ \bm{\hat{X}}_m $. Consistent with LNLN \citep{zhang2024towards}, we randomly erase between 0\% and 100\% of each modality sequence. Specifically, missing values in the visual and audio modalities are replaced with zeros, while missing tokens in the text modality are substituted with the \texttt{[UNK]} token used by BERT~\citep{devlin2019bert}.

\subsection{Text-aware Modality Enhancement}
Motivated by the informative nature of text in sentiment expression, we introduce the Text-aware Modality Enhancement (TME) module to standardize and enrich visual and audio features based on their semantic similarity to text and reconstruct corrupted or missing text semantics.

We first unify sequence length and feature dimension using a CTC-based temporal alignment \citep{zhou2024token} with a linear mapping: 
\begin{equation}
\{\bm{H}_v, \bm{H}_t, \bm{H}_a\} = \text{CTC}(\{\bm{\hat{X}}_{v}, \bm{\hat{X}}_{t},\bm{\hat{X}}_{a}\})
\end{equation}
where $\bm{H}_v, \bm{H}_t, \bm{H}_a \in \mathbb{R}^{L \times D}$ are the aligned features with length $L$ (set to $T_t$) and dimension $D$.
To enhance non-text modalities (i.e., visual), we compute token-wise similarities between visual and text tokens. Given L2-normalized token embeddings $\bm{v}_i$ and $\bm{t}_j$, their similarity score is calculated via temperature-scaled dot-product:
\begin{equation}
\bm{S}^{vt}_{i,j} = \frac{\exp(\langle \bm{v}_i, \bm{t}_j \rangle / \tau)}{\sum_{k=1}^{L} \exp(\langle \bm{v}_i, \bm{t}_k \rangle / \tau)}
\end{equation}
To suppress noisy or weakly correlated pairs, a hard threshold $\theta = 1/L$ is applied, generating a binary mask $\bm{M}^{vt}_{i,j} = \mathbb{I} [ \bm{S}^{vt}_{i,j} > \theta ]$.
The final enhanced visual representations are obtained as:
\begin{equation}
\bm{E}_v = \bm{H}_v + (\bm{M}^{vt} \odot \bm{S}^{vt}) \cdot \bm{H}_t
\end{equation}
The same way is applied to enhance the audio modality, resulting in enriched audio representations $\bm{E}_a$.
Given the rich sentiment semantics in text, we reconstruct missing textual information to enhance model robustness. A simple MLP-based decoder $\mathcal{R}(\cdot)$, consisting of two linear layers and a ReLU activation, is adopted as the reconstructor.
We employ Smooth L1 loss \citep{yuan2021transformer,sun2023efficient} to evaluate the text reconstruction:
\begin{equation}
\mathcal{L}_{\text{rec}} = \text{Smooth}_{\text{L}_1}(( {\bm{X}}_t - \mathcal{R}({\bm{H}}_t) ) \cdot (1 - \bm{P}_t) )
\end{equation}
where $\bm{P}_t$ is the text temporal mask indicating missing positions.
$\mathcal{L}_{\text{rec}}$ encourages the model to restore incomplete text semantics.
The reconstructed semantics in turn enhances non-text modalities.
\begin{figure}[t]
  \includegraphics[width=\linewidth]{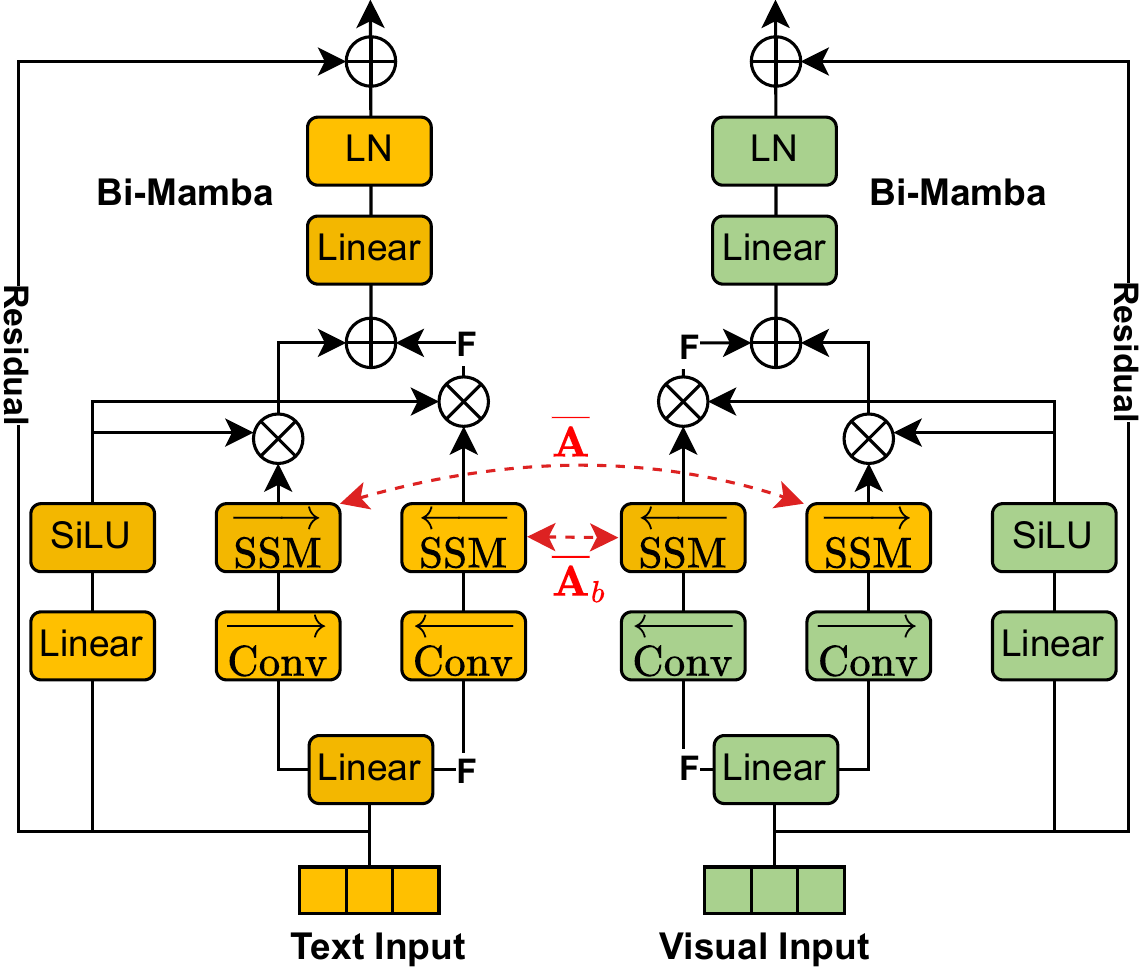}
\caption{An illustration of TC-Mamba with text and visual inputs. Red dashed lines indicate shared state transition matrices across Bi-Mamba blocks. The symbol F denotes the temporal flip operation.}
  \label{fig:TCMamba}
\end{figure}
\subsection{Text-based Context Mamba}
To capture text collaborative intra-modal contextual dependencies, we introduce the Text-based Context Mamba (TC-Mamba) to conduct efficient long-range modeling. 
TC-Mamba adopts bi-directional Mamba (Bi-Mamba) blocks~\citep{ZhuL0W0W24,yang2024cardiovascular} as the backbone, which learns bidirectional unimodal temporal patterns under the contextual supervision of text representations.

In Bi-Mamba, each modality stream employs forward and backward SSMs, parameterized by six matrices: $\overline{\mathbf{A}}, \overline{\mathbf{B}}, {\mathbf{C}}$ and $\overline{\mathbf{A}}_b, \overline{\mathbf{B}}_b, {\mathbf{C}}_b$. 
The transition matrices $\overline{\mathbf{A}}$ and $\overline{\mathbf{A}}_b$ govern temporal dynamics, while $\overline{\mathbf{B}}, \overline{\mathbf{B}}_b$ and ${\mathbf{C}}, {\mathbf{C}}_b$ manage the input and output operations. To capture temporal dependencies and common semantics, we share the bidirectional transition matrices $\overline{\mathbf{A}}$ and $\overline{\mathbf{A}}_b$ between the text modality and each non-text modality. 
In contrast, $\overline{\mathbf{B}}, \overline{\mathbf{B}}_b$ and ${\mathbf{C}}, {\mathbf{C}}_b$ remain independent to capture modality-specific information.

Figure~\ref{fig:TCMamba} illustrates the modeling process using text and visual inputs as an example. 
Specifically, the forward dynamics can be expressed as:
\begin{equation}
\bm{h}_t^{(t)} = \overline{\mathbf{A}} \, \bm{h}_{t-1}^{(t)} + \overline{\mathbf{B}^{t}} \, \bm{x}_t^{(t)}, \, \bm{y}_t^{(t)} = \mathbf{C}^{t} \, \bm{h}_t^{(t)}
\end{equation}
\begin{equation}
\bm{h}_t^{(v)} = \overline{\mathbf{A}} \, \bm{h}_{t-1}^{(v)} + \overline{\mathbf{B}^{v}} \, \bm{x}_t^{(v)}, \, \bm{y}_t^{(v)} = \mathbf{C}^{v} \, \bm{h}_t^{(v)}
\end{equation}
where $\bm{x}_t^{(t)}$, $\bm{x}_t^{(v)}$, $\bm{h}_t^{(t)}$, $\bm{h}_t^{(v)}$, and $\bm{y}_t^{(t)}$, $\bm{y}_t^{(v)}$ represent the input, hidden state, and output features at $t$ time step. The backward SSMs follows an analogous structure using the  $\overline{\mathbf{A}}_b$, $\overline{\mathbf{B}}_b$, and $\mathbf{C}_b$ matrices. The overall process of Bi-Mamba is formally as:
\begin{equation}
\left\{
\begin{aligned}
\bm{C}_t^1 &= \text{Bi-Mamba}(\bm{H}_t) \\
\bm{C}_v &= \text{Bi-Mamba}(\bm{E}_v)
\end{aligned}
\right.
\end{equation}
Similarly, the text-to-audio operation is denoted as:
\begin{equation}
\left\{
\begin{aligned}
\bm{C}_t^2 &= \text{Bi-Mamba}(\bm{H}_t) \\
\bm{C}_a &= \text{Bi-Mamba}(\bm{E}_a)
\end{aligned}
\right.
\end{equation}

In each TC-Mamba block, the text features are updated twice and we average them to obtain refined text representations $\bm{C}_t$:
\begin{equation}
\bm{C}_t = \text{Mean}(\bm{C}_t^1, \bm{C}_t^2)
\end{equation}
which capture the multimodal co-semantics and are used in reverse to iteratively update $\bm{C}_v$ and $\bm{C}_a$.

\subsection{Text-guided Query Mamba}
Building on the text-based context modeling in TC-Mamba, we further enhance cross-modal interactions with the proposed Text-guided Query Mamba (TQ-Mamba).
TQ-Mamba explicitly performs multimodal fusion by querying informative multimodal features and capturing cross-modal interactions between text and other modalities.

Specifically, it first leverages text-guided cross-attention to identify and query the most informative segments from the visual and audio streams. The query operation is formulated as:
\begin{equation}
\bm{Q}_f = \text{Cross-Attn}(\bm{C}_t, [\bm{C}_v; \bm{C}_a], [\bm{C}_v; \bm{C}_a])
\end{equation}
where $\bm{Q}_f$ represents the text-guided multimodal features, and $[\bm{C}_v; \bm{C}_a]$ denotes the concatenated sequence of visual and audio features. $\mathbf{Q}_f$ is then passed through latent Bi-Mamba blocks to learn intricate multimodal interactions: 
\begin{equation}
\bm{F}_z = \text{Bi-Mamba}(\bm{Q}_f)
\end{equation}
Finally, the fused feature $\bm{F}_z $ is aggregated using max pooling and projected via a fully connected layer (FC) to infer sentiment intensity.
\begin{equation}
\hat{\bm{Y}}  = \text{FC}(\text{MaxPool}(\bm{F}_z)),
\end{equation}
where $\hat{\bm{Y}}$ represents the predicted sentiment score.
\subsection{Overall Training Objective}
Our model is trained end-to-end with a combined loss that integrates sentiment prediction and text reconstruction objectives. 
The sentiment prediction loss $\mathcal{L}_{\text{task}}$ can be described as:
\begin{equation}
    \mathcal{L}_{\text{task}} = \frac{1}{N_b} \sum_{n=1}^{N_b} \| \bm{Y}^n - \hat{\bm{Y}}^n \|_2^2
\end{equation}
Therefore, the overall loss $\mathcal{L} $ can be written as:
\begin{equation}
    \mathcal{L} = \mathcal{L}_{\text{task}} + \lambda \mathcal{L}_{\text{rec}},
\end{equation}
where $\mathcal{L}_{\text{rec}}$ is the reconstruct loss mentioned above and hyperparameter $\lambda$  balances the two terms.
\begin{table}[h]
\centering
\resizebox{\linewidth}{!}{
\begin{tabular}{cccccc}
\toprule
{Dataset} & {\#Train} & {\#Valid} & {\#Test} & {\#Total} & {Language} \\
\midrule
MOSI       & 1284  & 229  & 686  & 2199  & English \\
MOSEI      & 16326 & 1871 & 4659 & 22856 & English \\
SIMS    & 1368  & 456  & 457  & 2281  & Chinese \\
\bottomrule
\end{tabular}
}
\caption{The statistics of MOSI, MOSEI, and SIMS.}
\label{tab:datasets}
\end{table}

\begin{table}[htbp]
\centering
\small
\begin{tabular}{ccccc}  
\toprule
 Descriptions & {MOSI} & {MOSEI} & {SIMS} \\
\midrule
Length $L$ & 50 & 50 & 39 \\
Mamba State  & 12 & 12 & 16 \\ 
Mamba Expansion & 4 & 4 & 2 \\ 
Mamba Depth & \{1,1\} & \{2,2\} & \{1,2\} \\
Attention Head & 8 & 8 & 8 \\ 
Loss Weight $\lambda$ & 0.7 & 0.3 & 1.0\\
Warm Up & \checkmark & \checkmark & \checkmark \\
\bottomrule
\end{tabular}
\caption{Hyper-parameters settings on different datasets.}
\label{tab:Hyperparameters}
\end{table}

\begin{table*}[t]
\centering
\resizebox{\textwidth}{!}{ 
\begin{tabular}{ccccccccccccc}
\toprule
\multirow{2}{*}{Method} & \multicolumn{6}{c}{MOSI} & \multicolumn{6}{c}{MOSEI} \\
\cmidrule(lr){2-7} \cmidrule(lr){8-13}
 & Acc-7 & Acc-5 & Acc-2 & F1 & MAE & Corr & Acc-7 & Acc-5 & Acc-2 & F1 & MAE & Corr \\
\midrule
MISA & 29.85 & 33.08 & 71.49 / 70.33 & 71.28 / 70.00 & 1.085 & 0.524 & 40.84 & 39.39 & 71.27 / 75.82 & 63.85 / 68.73 & 0.780 & 0.503 \\
Self-MM & 29.55 & 34.67 & 70.51 / 69.26 & 66.60 / 67.54 & 1.070 & 0.512 & 44.70 & 45.38 & 73.89 / 77.42 & 68.92 / 72.31 & 0.695 & 0.498 \\
MMIM & 31.30 & 33.77 & 69.14 / 67.06 & 66.65 / 64.04 & 1.077 & 0.507 & 40.75 & 41.74 & 73.32 / 75.89 & 68.72 / 70.32 & 0.739 & 0.489 \\
TFR-Net & 29.54 & 34.67 & 68.15 / 66.35 & 61.73 / 60.06 & 1.200 & 0.459 & 46.83 & 34.67 & 73.62 / 77.23 & 68.80 / 71.99 & 0.697 & 0.489 \\
CENET & 30.38 & 33.62 & 71.46 / 67.73 & 68.41 / 64.85 & 1.080 & 0.504 & \textbf{47.18} & \textbf{47.83} & 74.67 / 77.34 & 70.68 / 74.08 & 0.685 & 0.535 \\
ALMT & 30.30 & 33.42 & 70.40 / 68.39 & 72.57 / 71.80 & 1.083 & 0.498 & 40.92 & 41.64 & 76.64 / 77.54 & 77.14 / 78.03 & 0.674 & 0.481 \\
BI-Mamba & 31.20 & 34.02 & 71.74 / 71.12 & 71.83 / 71.11 & 1.087 & 0.498 & 45.12 & 45.76 & 76.82 / 76.72 & 76.35 / 76.38 & 0.701 & 0.545 \\
LNLN & 32.53 & 36.25 & 71.91 / 70.11 & 71.71 / 70.02 & 1.062 & 0.503 & 45.42 & 46.17 & 76.30 / \textbf{78.19} & \textbf{77.77} / \textbf{79.95} & 0.692 & 0.530 \\
\midrule
\textbf{TF-Mamba} & \textbf{33.95} & \textbf{37.74} & \textbf{73.46} / \textbf{72.54} & \textbf{73.59} / \textbf{72.57} & \textbf{1.035} & \textbf{0.548} 
& 45.66 & \underline{46.64} & \textbf{77.34} / \underline{77.61} & \underline{77.18} / 77.43 & \textbf{0.673} & \textbf{0.578} \\
\bottomrule
\end{tabular}%
}
\caption{Overall performance comparison on the MOSI and MOSEI datasets under missing modality settings.}
\label{tab:robustness_comparison}
\end{table*}
\begin{figure*}[t]
    \centering
    \includegraphics[width=\linewidth]{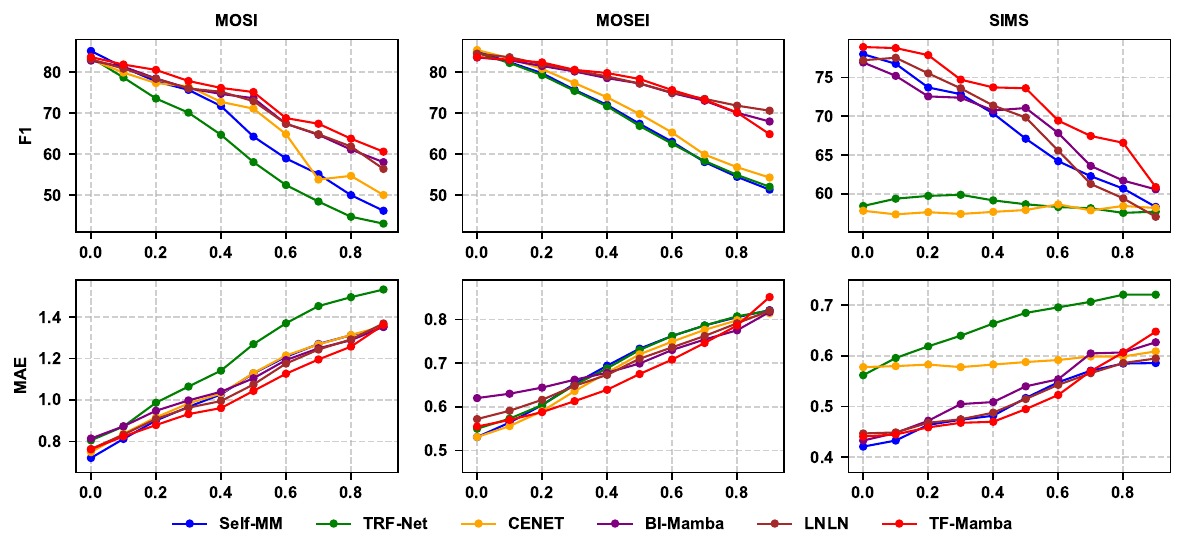}
    \caption{Performance trends of models under varying missing rates on MOSI, MOSEI, and SIMS datasets.}
    \label{fig:enter-label}
\end{figure*}

\section{Experiments}
\subsection{Datasets and Evaluation Metrics}
We evaluate our method on three widely used MSA datasets: {MOSI}~\cite{zadeh2016multimodal}, {MOSEI}~\cite{zadeh2018multimodal}, and {SIMS}~\cite{yu2020ch}. All experiments are conducted under the unaligned data setting.  We adopt the publicly released features provided by each benchmark. Table \ref{tab:datasets} shows the statistic details. 
The details of feature extraction procedures are reported in Appendix~\ref{sec:feature-extraction}.

For evaluation, following \citet{zhang2024towards}, we report 5-class (Acc-5) and 7-class (Acc-7) Accuracy on MOSI and MOSEI, and 3-class (Acc-3) and 5-class (Acc-5) Accuracy on SIMS. Additionally, 2-class Accuracy (Acc-2), mean absolute error (MAE), Pearson correlation (Corr), and F1 score (F1) are reported on all datasets.
For Acc-2 and F1 on MOSI and MOSEI, results are provided under two settings: negative vs. positive (left of "/") and negative vs. non-negative (right of "/"). Except for MAE, higher values indicate better performance.

\subsection{Implementation Details}
To ensure fair comparison and evaluation, we follow the same missing modality setting from LNLN \citep{zhang2024towards}, where training data is randomly dropped with an uncertain probability. During testing, the missing rate $r$ is varied from 0 to 0.9 in increments of 0.1. 
Evaluation at $r=1.0$ is excluded as it removes all modalities, making the task ill-posed. 
Final results are averaged across all missing rates to assess model robustness.

All experiments are conducted using PyTorch 2.1.0. Models are optimized with AdamW and a cosine annealing learning rate schedule, starting at 0.0001. The feature dimension for each modality is unified to 128.
Training is performed for 200 epochs with a batch size of 64 on a single NVIDIA Tesla V100-DGXS GPU (32 GB memory).
More implementation details are provided in Table \ref{tab:Hyperparameters}.

\subsection{Baseline Models}
To  evaluate the robustness of TF-Mamba, we conduct fair comparisons with a range of state-of-the-art (SOTA) methods under the same missing modality conditions. The compared baselines include  MISA~\cite{hazarika2020misa}, Self-MM~\cite{yu2021learning}, MMIM~\cite{han2021improving}, CENET~\cite{wang2022cross}, TETFN~\cite{wang2023tetfn}, TFR-Net~\cite{yuan2021transformer}, ALMT~\cite{zhang2023learning}, BI-Mamba~\cite{yang2024cardiovascular}, and LNLN~\cite{zhang2024towards}. 
Detailed baseline settings and descriptions are provided in Appendix~\ref{sec:baseline-comparisons}.

\begin{table}[t]
\centering
\resizebox{\linewidth}{!}{ 
\begin{tabular}{ccccccc}
\toprule
Method & Acc-5 & Acc-3 & Acc-2 & F1 & MAE & Corr \\
\midrule
MISA & 31.53 & \textbf{56.87} & 72.71 & 66.30 & 0.539 & 0.348 \\
Self-MM & 32.28 & 56.75 & 72.81 & 68.43 & \textbf{0.508} & 0.376 \\
MMIM & 31.81 & 52.76 & 69.86 & 66.21 & 0.544 & 0.339 \\
TFR-Net & 26.52 & 52.89 & 68.13 & 58.70 & 0.661 & 0.169 \\
CENET & 22.29 & 53.17 & 68.13 & 57.90 & 0.589 & 0.107 \\
ALMT & 20.00 & 45.36 & 69.66 & \textbf{72.76} & 0.561 & 0.364 \\
BI-Mamba & 31.90 & 54.95 & 70.79 & {69.26} & 0.529 & 0.345 \\
LNLN & 33.08 & 56.01 & {73.62} & 68.84 & 0.514 & \textbf{0.389} \\
\midrule
\textbf{TF-Mamba} & \textbf{34.46} & {55.51} & \textbf{74.68} & \underline{72.20} & \underline{0.512 }& \underline{0.386} \\
\bottomrule
\end{tabular}%
}
\caption{Overall performance comparison on the SIMS dataset under missing modality settings.}
\label{tab:sims_performance}
\end{table}

\begin{table}[t]
\centering
\resizebox{\linewidth}{!}{
\begin{tabular}{ccccccccc}
\toprule
\multirow{2}{*}{\textbf{Model}} & \multicolumn{3}{c}{{MOSI}} & \multicolumn{3}{c}{{SIMS}} \\
\cmidrule(lr){2-4} \cmidrule(lr){5-7}
 & MAE & F1 & Acc-7 & MAE & F1 & Acc-5 \\
\midrule
w/o  Enhancement & 1.104 & 72.78 & 29.17 & 0.522 & 68.83 & 29.67 \\
w/o  Reconstruction & 1.085 & 73.09 & 31.73 & 0.520 & 70.21 & 32.89\\
\midrule
w/o TME & 1.071 & 73.06 & 32.81 & 0.517 & 69.07 & 33.24 \\
TME + BI-Mamba & 1.059 & \textbf{74.17} & 32.52 & \textbf{0.512} & 70.20 & 32.58 \\
\midrule
w/o TC-Mamba & 1.064 & 73.23 & 32.39 & 0.517 & 68.33 & 34.35 \\
w/o TQ-Mamba & 1.055 & 72.68 & 32.55 & 0.517 & 69.44 & 33.79 \\
\midrule
\textbf{TF-Mamba} & \textbf{1.035} & {73.46} & \textbf{33.95} & \textbf{0.512} & \textbf{72.20} & \textbf{34.46} \\
\bottomrule
\end{tabular}
}
\caption{Ablation study of different modules and strategies on the MOSI and SIMS datasets.}
\label{tab:ablation_study}
\end{table}
\subsection{Robustness Comparison}
Tables~\ref{tab:robustness_comparison} and~\ref{tab:sims_performance} present the robustness results on the MOSI, MOSEI, and SIMS datasets. The best results are shown in bold, while the second-best results (ours) are underlined.
{TF-Mamba} achieves superior performance across most metrics. Compared with Transformer-based models (CENET, TETFN, TFR-Net, and ALMT), Mamba-based approaches gain performance, underscoring their potential as competitive alternatives.
Notably, TF-Mamba outperforms BI-Mamba, owing to its text enhancement strategy. 
On MOSI, TF-Mamba outperforms the previous SOTA model LNLN by {4.36\%} in Acc-7 and {2.62\%} in F1 score, showing its strong resilience to varying missing rates. 
On MOSEI, LNLN achieves better binary classification results, likely benefiting from the larger data scale.
TF-Mamba also achieves strong performance on SIMS, obtaining the best five-class score of 34.46\% and binary classification Accuracy of 74.68\%.

Additionally, as shown in Figure~\ref{fig:enter-label}, all models degrade as the missing rate increases, reflecting their sensitivity to incomplete data (see Appendix \ref{sec:Robust Details} for more details).
TF-Mamba still maintains relatively stable performance under varying missing rates.
These results validate the robustness of TF-Mamba and its promise for robust MSA.

\subsection{Ablation Study}
We conduct ablation experiments to assess the effectiveness of the strategies and modules within TF-Mamba. The average results on the MOSI and SIMS datasets are summarized in Table~\ref{tab:ablation_study}.

\paragraph{Effect of Text Modality}
When the text enhancement strategy is removed, model performance degrades notably, which underscores the effectiveness of text-enhanced fusion for sentiment prediction. 
Similarly, without the text reconstruction loss, the model struggles to recover missing textual semantics, leading to poorer multimodal representations. This highlights the importance of reconstructing incomplete text features for robust MSA.
We further evaluate model performance under complete modality missing conditions ($r=1.0$), with results and discussions summarized in Appendix \ref{sec:modality_missing}. 
The complete absence of the text modality causes a marked performance drop, even when combined with missing audio or visual inputs. In contrast, missing only audio or visual modality has a limited impact. 
These findings further underscore the significance of text enhancement in robust MSA.

\paragraph{Module Design of TME }
Removing the TME module leads to consistent performance drops on both datasets. 
These results show that TME effectively facilitates representation learning and multimodal fusion through text-aware alignment and enhancement.  
In addition, incorporating the TME module into the baseline BI-Mamba improves Acc-7 by 4.23\% on MOSI and Acc-5 by 2.13\% on SIMS.
Such outcomes substantiate both the effectiveness and rationality of the TME module.
\paragraph{Role of TC-Mamba and TQ-Mamba}
Eliminating either TC-Mamba or TQ-Mamba leads to performance drops. When TC-Mamba is removed, the model struggles to learn text-related context within non-text modalities, which reduces overall effectiveness and underscores the role of TC-Mamba in collaborative intra-modal modeling. 
The impact is even greater when TQ-Mamba is excluded, indicating that guiding multimodal fusion with text queries is vital for producing more robust and expressive joint representations. Together, these modules strengthen both intra-modal representation learning and cross-modal fusion, resulting in improved robustness and predictive accuracy.

\subsection{Efficiency Study}
We conduct an efficiency analysis of TF-Mamba, focusing on its intra-modal modeling and cross-modal interaction components.
To ensure fair comparisons, we exclude the computational cost of pre-trained encoders.
All experiments are performed under identical  conditions, evaluating model parameters (Params), floating-point operations (FLOPs), and performance (Acc-7 and F1) on  MOSI dataset. 
Comparisons are made against the Transformer-based SOTA baseline LNLN and a TF-Mamba variant, TF-Trans, where Mamba blocks are replaced with Transformer blocks.

\begin{figure}[t]
    \centering
    \includegraphics[width=\linewidth]{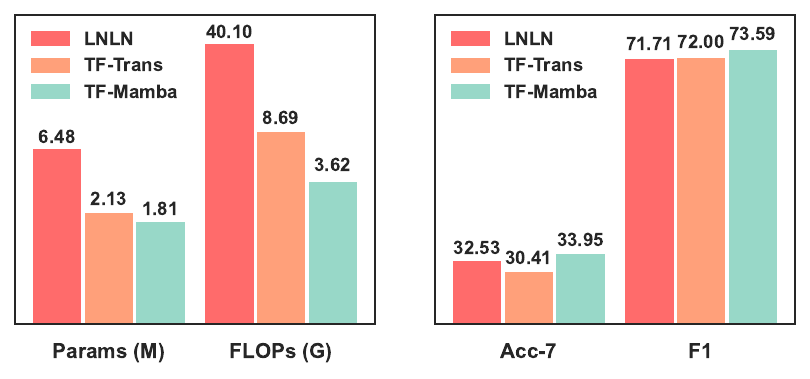}
    \caption{Model performance and complexity comparison during inference on MOSI dataset.}
    \label{fig:efficiency}
\end{figure}

As shown in Figure~\ref{fig:efficiency}, TF-Mamba achieves superior performance while significantly improving efficiency over previous Transformer-based fusion models. 
Benefiting from the linear complexity of Mamba, TF-Mamba requires fewer parameters and reduces inference costs by 36.48G FLOPs compared to LNLN.
Although replacing Mamba with Transformers in TF-Trans increases computational overhead, it retains competitive robustness and Accuracy.
These results demonstrate that TF-Mamba offers substantial computational advantages while maintaining strong performance, validating the effectiveness of its efficient modeling and fusion strategy for robust MSA with missing modalities.

\begin{figure}[t]
    \centering
    \includegraphics[width=\linewidth]{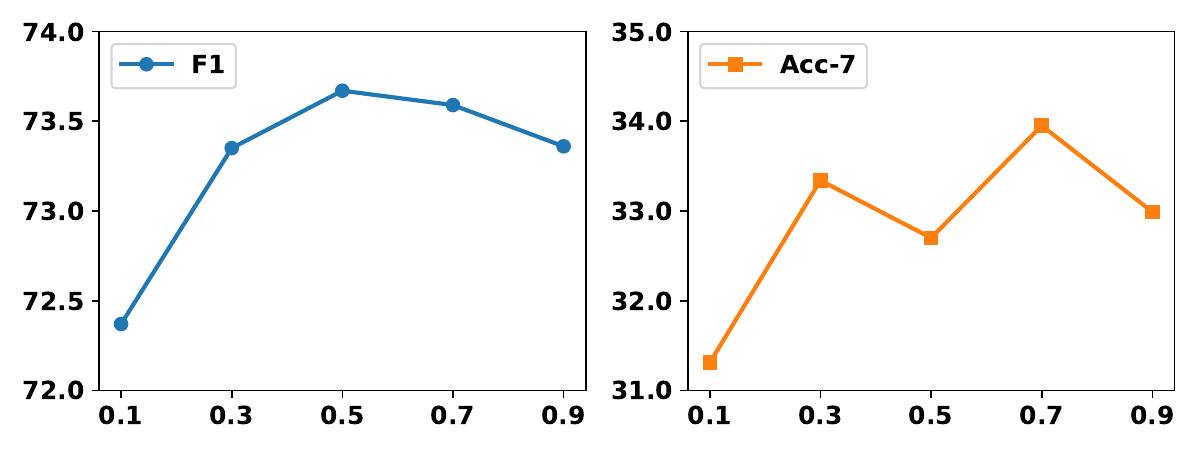}
\caption{Effect of the regularization parameter $\lambda$ on F1 Score and Acc-7 on the MOSI dataset.}   
\label{fig:lambda_effect}
\end{figure}

\begin{table}[t]
\centering
\resizebox{0.95\linewidth}{!}{
\begin{tabular}{ccccc}
\toprule
TC-Mamba & TQ-Mamba & MAE & F1 & Acc-5 \\
\midrule
1 & 1 & 0.521 & 71.18 & 33.83 \\
1 & 2 & 0.512 & \textbf{72.20} & \textbf{34.46} \\
1 & 3 & 0.523 & 69.36 & 34.31 \\
2 & 1 & 0.522 & 70.46 & 33.04 \\
2 & 2 & \textbf{0.508} & 70.97 & 34.09 \\
2 & 3 & 0.518 & 70.15 & 33.85 \\
\bottomrule
\end{tabular}
}
\caption{Different Mamba layer settings and model performance on SIMS dataset.}
\label{tab:mamba_performance_sims}
\end{table}
\subsection{Sensitivity Analysis}
\label{sec:sensitivity_analysis}
In this section, we analyze the sensitivity of key hyperparameters on model performance.  We focus on two factors: (1) the reconstruction loss weight, controlled by $\lambda$, and (2) the number of layers in the TC-Mamba and TQ-Mamba modules, which govern model capacity and computational cost.

\paragraph{Effect of Loss Weight $\lambda$}  
We first examine the influence of the reconstruction loss weight $\lambda$ on model performance. This hyperparameter controls the relative importance of the reconstruction loss in guiding the model to recover missing semantic information. We evaluated the model across $\lambda \in \{0.1, 0.3, 0.5, 0.7, 0.9\}$ on the MOSI dataset, and the results are visualized in Figure~\ref{fig:lambda_effect}.  
The analysis reveals a clear trend: excessively small values (e.g., 0.1) provide insufficient supervision for reconstructing missing semantics, leading to suboptimal sentiment prediction. Conversely, excessively large values (e.g., 0.9) cause the reconstruction loss to dominate, which interferes with the model's primary sentiment inference. A gradual increase in $\lambda$ leads to consistent gains in F1 and Acc-7, with the overall best performance observed at $\lambda = 0.7$. This choice strikes an effective balance between reconstruction and sentiment prediction, underscoring the necessity of tuning $\lambda$ appropriately.

\paragraph{Effect of Mamba Layers}
We further investigate the impact of varying the number of TC-Mamba and TQ-Mamba layers on model performance. The evaluation on the SIMS dataset is summarized in Table~\ref{tab:mamba_performance_sims}. The results show that using too few layers fails to capture the complex cross-modal interactions, while excessive layers introduce higher computational cost without consistent performance gains and may even risk overfitting. Overall, performance improves as the number of layers increases, with the (1,2) configuration of TC-Mamba and TQ-Mamba achieving the best trade-off between efficiency and Accuracy, delivering strong results at a moderate computational cost.

\subsection{Further Analysis}
\begin{figure}[t]
    \centering
    \includegraphics[width=\linewidth]{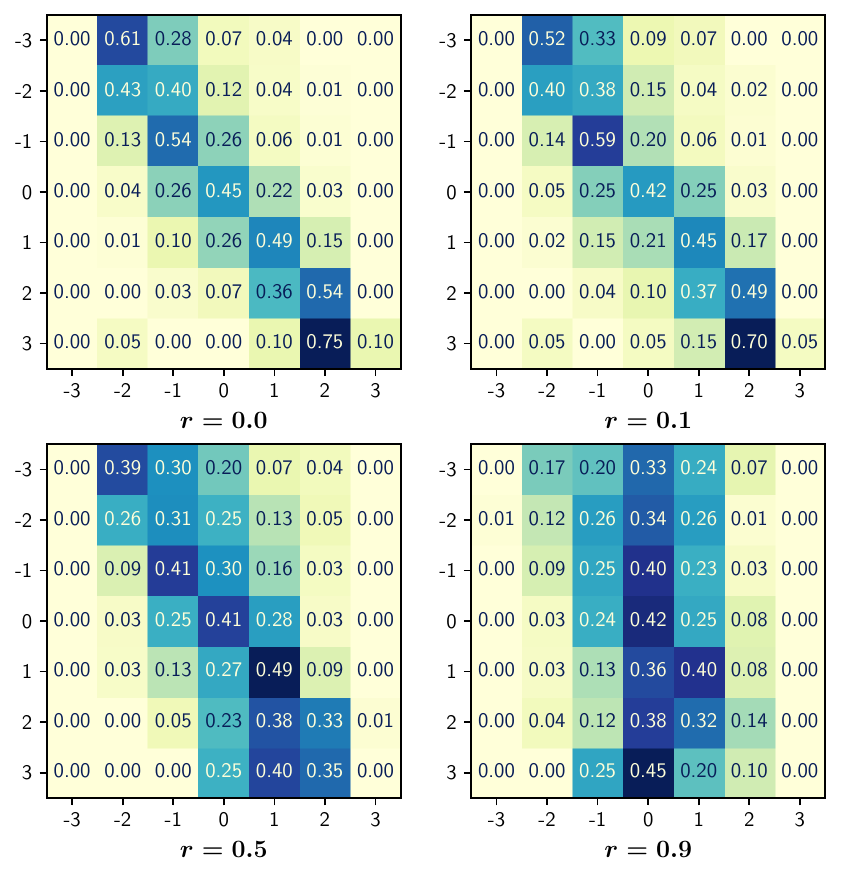}
    \caption{Seven-class confusion matrices of TF-Mamba on the MOSI dataset.
Labels -3 to 3 represent sentiment levels from strongly negative to strongly positive.}
    \label{fig:confusion_matrix}
\end{figure}

To further assess the effectiveness and robustness of our approach, we visualize the confusion matrices on the MOSI dataset under different missing rates in Figure~\ref{fig:confusion_matrix}. As expected, the model’s robustness and performance progressively decline as the missing rate $r$ increases. At $r=0$, the model achieves strong classification performance, with high diagonal values indicating effective category discrimination. When the missing rate increases to $r=0.5$, classification Accuracy drops due to partial data loss, though the diagonal values remain relatively high, suggesting the model still correctly classifies most samples.
At a severe missing rate of $r=0.9$, excessive data missing leads the model to favor certain categories, exhibiting the typical ``lazy'' prediction behavior described in LNLN.
Nevertheless, our model does not degenerate into random guessing, retaining a learned bias towards sentiment-relevant categories.
These findings demonstrate the robustness of TF-Mamba in handling varying levels of incomplete data and underscore the importance of designing tailored fusion strategies for modeling incomplete data.
\section{Conclusion}
In this paper, a novel and efficient text-enhanced fusion Mamba (TF-Mamba) framework is developed to tackle
random missing modality issues in the MSA task. 
By integrating text enhancement strategies into the Mamba-based fusion architecture, our model effectively captures cross-modal interactions and informative multimodal representations with low computational overhead.
Extensive experiments on three public benchmarks demonstrate that TF-Mamba outperforms current leading baselines across varying levels of data incompleteness. 
Further analysis reveals that TF-Mamba achieves notable reductions in parameters and computational costs relative to Transformer-based methods, underscoring the advantages and potential of Mamba-based fusion techniques in achieving robust MSA.
\section*{Limitations}
Although TF-Mamba demonstrates strong robustness and efficiency under various missing modality scenarios, it has yet to be evaluated under more complex real-world missing patterns. 
The proposed method may experience minor performance degradation when applied to real-world scenarios. In addition, our work relies on pre-extracted unimodal features, which may limit its end-to-end optimization potential.  In the future, we will explore more complex modality missing cases and develop suitable methods to overcome these limitations.
\section*{Acknowledgments}
This work was supported by the National Natural Science Foundation of China (Nos. 62272025 and U22B2021), and by the Fund of the State Key Laboratory of Software Development Environment.

\bibliography{custom}

\appendix
\section{Feature Extraction} \label{sec:feature-extraction}
To ensure fair and consistent comparisons with existing SOTA methods, we adopt the officially released features provided by the corresponding benchmark datasets, following the unified MMSA framework \citep{mao2022m}. All features are pre-extracted for text, audio, and visual modalities using standard toolkits, as detailed below.
\paragraph{Text Modality} 
For {MOSI} and {MOSEI}, the {bert-base-uncased} \citep{devlin2019bert} model is used to encode raw utterances. For {SIMS}, the {bert-base-chinese} model is employed. The extracted text feature dimension is 768 for all datasets, with input sequence lengths of 50, 50, and 39 for MOSI, MOSEI, and SIMS, respectively.
\paragraph{Audio Modality}
For {MOSI} and {MOSEI}, COVAREP \citep{degottex2014covarep} is applied to extract 5 and 74 low-level acoustic features, including pitch, glottal source parameters, and MFCCs, with input sequence lengths of 375 and 500, respectively. For {SIMS}, Librosa \citep{mcfee2015librosa} is used to extract 33-dimensional audio features with a sequence length of 400.
\paragraph{Visual Modality}
For {MOSI} and {MOSEI}, Facet \citep{imotions2017fea} is utilized to extract 20 and 35 facial-related features, such as facial action units and head pose, with sequence lengths of 500. For {SIMS}, OpenFace 2.0 \citep{8373812} is used to extract 709-dimensional visual features with a sequence length of 55.
\paragraph{Sentiment Labels}
For {MOSI} and {MOSEI}, sentiment labels range from $-3$ to $3$, representing sentiment intensity from strongly negative to strongly positive. For {SIMS}, labels are scaled within $-1$ to $1$ with the same polarity interpretation.
\section{Baseline Settings} \label{sec:baseline-comparisons}
We compare TF-Mamba against several MSA baselines. All experiments are conducted under the same dataset settings for fairness. We reproduce BI-Mamba and LNLN in our environment, while the results for other baselines are reported from~\cite{zhang2024towards}. Below are brief descriptions of them.

    \textbf{MISA}~\cite{hazarika2020misa} models both shared and private features across modalities to improve robustness in sentiment prediction.
    
    \textbf{Self-MM}~\cite{yu2021learning} generates unimodal labels and conducts multi-task training to capture both consistent and differential representations.
    
    \textbf{MMIM}~\cite{han2021improving} introduces mutual information maximization techniques for multimodal fusion to better capture correlated representations between modalities.

    \textbf{CENET}~\cite{wang2022cross} is a cross-modal enhancement network that adaptively aligns and fuses multimodal signals through cross-attentive interaction mechanisms.

    \textbf{TFR-Net}~\cite{yuan2021transformer} is a Transformer-based model incorporating a fusion-reconstruction strategy to enhance sentiment prediction performance under incomplete modality conditions.

    \textbf{ALMT}~\cite{zhang2023learning} utilizes language features to suppress irrelevant or conflicting information from visual and audio inputs and learn complementary multimodal representations.

    \textbf{BI-Mamba}~\cite{yang2024cardiovascular} integrates bidirectional Mamba networks for multi-view medical image fusion. In our adaptation, we treat multimodal features as multi-view inputs to suit the model.

    \textbf{LNLN}~\cite{zhang2024towards} treats the language modality as dominant and introduces a dominant modality correction and dominant modality-based multimodal learning to enhance robustness against noisy and missing modality scenarios.
\section{Details of Robust Comparison} \label{sec:Robust Details}
Tables~\ref{tab:Details_Robust_Compare_MOSI}, \ref{tab:Details_Robust_Compare_MOSEI}, and \ref{tab:Details_Robust_Compare_SIMS} present detailed robustness comparisons on the MOSI, MOSEI, and SIMS datasets, respectively. We observe that when the missing rate $r$ is low, Self-MM achieves notable advantages across several evaluation metrics. However, as $r$ increases, TF-Mamba consistently outperforms other methods on most metrics, demonstrating its ability to learn robust multimodal representations under varying levels of data incompleteness. Additionally, while models such as LNLN and TF-Mamba perform well under high missing modality rates, they often struggle to maintain optimal performance when modality missingness is low. Balancing robustness and Accuracy across different missing conditions remains challenging.

\section{Analysis of Complete Modality Missing}
\label{sec:modality_missing}
We conduct comprehensive experiments under complete modality missing conditions to further assess model robustness and examine the impact of discarding different modalities. The detailed results on the MOSI, MOSEI, and SIMS datasets are presented in Tables~\ref{tab:Missing MOSI}, \ref{tab:Missing MOSEI}, and \ref{tab:Missing SIMS}, respectively.

These results consistently show that missing the text modality (T) leads to the largest performance drop, whether alone or combined with the audio (A) or visual (V) modality.  This reveals the central role of text modality in MSA,  as it typically provides the most direct and sentiment-rich information. In contrast, audio and visual modalities offer complementary cues, with their absence causing relatively minor degradation. This can be attributed to our text enhancement strategy, which provides effective task and sentiment information when other modalities are missing, thereby maintaining robust performance. 
Importantly, models trained under random missing assumptions struggle with complete modality loss, highlighting the need for customized fusion strategies to better handle structured or extreme missing scenarios.
\begin{table}[ht]
    \centering
    \resizebox{\linewidth}{!}{
    \begin{tabular}{ccccccc}
        \toprule   
        \multirow{2}{*}{Missing Condition} & \multicolumn{6}{c}{MOSI} \\
        \cmidrule(lr){2-7}
        & Acc-7 & Acc-5 & Acc-2 & F1 & MAE & Corr \\ 
        \midrule
        Missing T & 19.39 & 19.97 & 52.59 / 53.79 & 51.33 / 52.31 & 1.505 & 0.108 \\
        Missing A  & 42.71 & 49.71 & 83.84 / 81.92 & 83.88 / 81.90 & 0.782 & 0.765 \\
        Missing V  & 42.27 & 48.40 & 82.32 / 81.05 & 82.42 / 81.10 & 0.778 & 0.772 \\
        Missing T \& A & 16.03 & 16.76 & 57.01 / 55.83 & 54.41 / 53.10 & 1.446 & 0.038 \\
        Missing T \& V & 16.91 & 19.10 & 55.49 / 56.27 & 54.86 / 55.45 & 1.590 & 0.183 \\
        Missing A \& V  & 41.40 & 47.52 & 84.60 / 82.22 & 84.54 / 82.07 & 0.783 & 0.767 \\
        \midrule
        {Average}  & {29.78} & {33.58} & {69.31 / 68.51} & {68.57 / 67.65} & {1.147} & {0.439} \\

        {TF-Mamba} & {33.95} & {37.74} & {73.46} / {72.54} & {73.59} / {72.57} & {1.035} & {0.548} \\
        \bottomrule
    \end{tabular}
    }
    \caption{Performance of TF-Mamba under complete modality missing settings on MOSI dataset.}
    \label{tab:Missing MOSI}

\end{table}

\begin{table}[ht]
    \centering
    \resizebox{\linewidth}{!}{
    \begin{tabular}{ccccccc}
        \toprule   
        \multirow{2}{*}{Method} & \multicolumn{6}{c}{MOSEI} \\
        \cmidrule(lr){2-7}
        & Acc-7 & Acc-5 & Acc-2 & F1 & MAE & Corr \\ 

        \midrule
        Missing T & 36.40 & 36.40 & 60.68 / 66.52 & 61.16 / 63.94 & 0.943 & 0.152 \\
        Missing A  & 49.58 & 50.78 & 83.32 / 81.99 & 83.16 / 82.02 & 0.569 & 0.742 \\
        Missing V  & 50.44 & 52.11 & 83.13 / 83.30 & 82.71 / 83.15 & 0.565 & 0.738 \\
        Missing T \& A & 37.39 & 37.39 & 61.81 / 69.14 & 60.27 / 62.93 & 0.886 & 0.142 \\
        Missing T \& V  & {41.36} & {41.36} & 62.85 / {71.02} & 48.51 / 58.99 & 0.855 & 0.124 \\
        Missing A \& V & 50.01 & 51.43 & 83.68 / 78.06 & 83.61 / 78.87 & 0.608 & 0.737 \\
        \midrule
        {Average}  & {44.20} & {44.91} & {72.59 / 75.00} & {69.90 / 71.65} & {0.738} & {0.439} \\

         {TF-Mamba} & 45.66 & {46.64} & {77.34} / 77.61 & {77.18} / 77.43 & {0.673} & {0.578} \\
        \bottomrule
    \end{tabular}
    }
    \caption{Performance of TF-Mamba under complete modality missing settings on MOSEI dataset.}
    \label{tab:Missing MOSEI}
\end{table}

\begin{table}[ht]
    \centering
    \small
    \resizebox{\linewidth}{!}{
    \begin{tabular}{ccccccc}
        \toprule   
        \multirow{2}{*}{Method} & \multicolumn{6}{c}{SIMS} \\
        \cmidrule(lr){2-7}
        & Acc-5 & Acc-3 & Acc-2 & F1 & MAE & Corr \\ 
        \midrule
        Missing T & {26.91} & 49.67 & 69.37 & 56.82 & 0.738 & 0.043 \\
        Missing A & 35.01 & 61.71 & 73.96 & 74.18 & 0.456 & 0.535 \\
        Missing V  & 36.76 & 62.58 & 77.68 & 77.10 & 0.447 & 0.539 \\
        Missing T \& A & 20.79 & 34.79 & 68.71 & 56.50 & 0.759 & 0.040  \\
        Missing T \& V   & 18.60 & 30.63 & 36.76 & 32.93 & 0.907 & 0.014 \\
        Missing A \& V & 35.23 & 61.27 & {78.56} & {77.33} & 0.459 & 0.532 \\
        \midrule
        {Average}  & {28.88} & {50.11} & {67.51} & {62.48} & {0.628} & {0.284} \\
        {TF-Mamba} & {34.46} & {55.51} & {74.68} & {72.20} & {0.512 }& {0.386} \\
        \bottomrule
    \end{tabular}
    }
    \caption{Performance of TF-Mamba under complete modality missing settings on SIMS dataset.}
    \label{tab:Missing SIMS}
\end{table}

\begin{figure}[t]
    \centering
    \includegraphics[width=\linewidth]{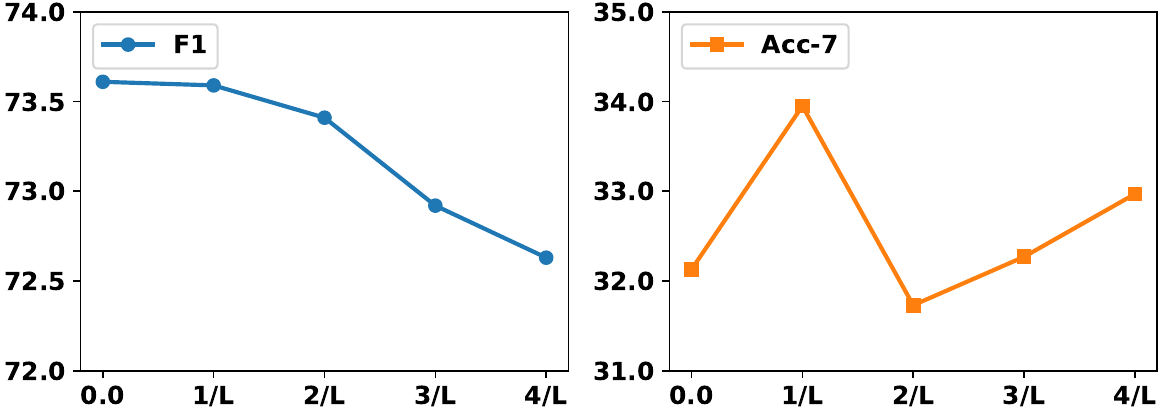}
    \caption{Model performance: F1 and Acc-7 on the MOSI dataset under different threshold settings.}
    \label{fig:mosi_threshold}
\end{figure}

\begin{figure}[t]
    \centering
    \includegraphics[width=\linewidth]{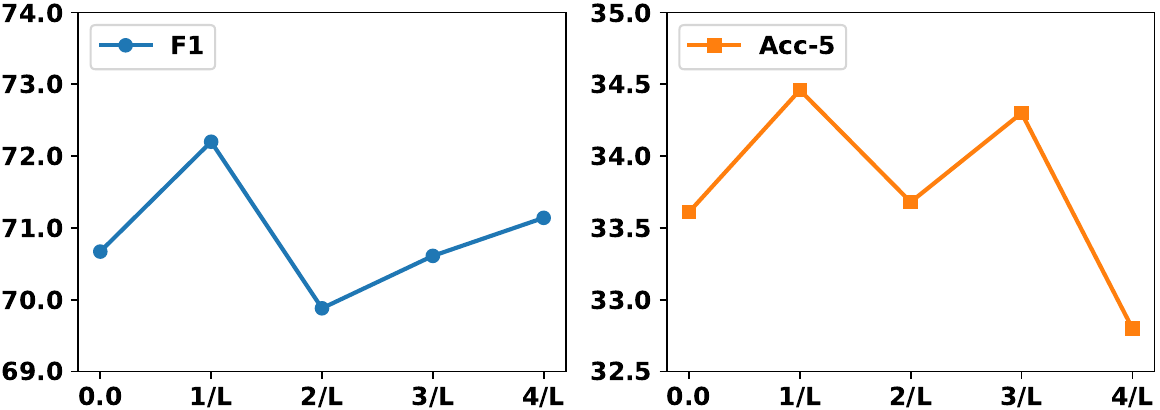}
    \caption{Model performance: F1 and Acc-5 on the SIMS dataset under different threshold settings.}
    \label{fig:sims_threshold}
\end{figure}
\section{Analysis of Threshold Selection in TME}
In the TME module, a threshold $\theta$ is introduced to generate binary masks over the similarity matrix between text tokens and other modalities. This mechanism suppresses irrelevant information by filtering out weakly correlated token pairs. The threshold is set as the average similarity value $1/L$, where $L$ denotes the number of text tokens. 
Token pairs with similarity scores below the threshold are masked, whereas those exceeding the threshold are retained, thereby mitigating the influence of weak correlations while ensuring the preservation of informative semantic relationships.

To assess robustness, we conduct a sensitivity analysis by varying the threshold within ${0, 1/L, 2/L, 3/L, 4/L}$ on the MOSI and SIMS datasets. The corresponding model performance is shown in Figure~\ref{fig:mosi_threshold} and Figure~\ref{fig:sims_threshold}.
From Figure~\ref{fig:mosi_threshold} (MOSI), a threshold of $1/L$ provides the overall best performance across F1 and Acc-7, whereas $\theta=0$ retains excessive irrelevant information and higher thresholds gradually remove important cues.
Figure~\ref{fig:sims_threshold} (SIMS) exhibits a similar trend: $1/L$ consistently yields stable and optimal performance across both F1 and Acc-5, while $\theta=0$ degrades performance and overly high thresholds suppress too much semantic information.
Overall, this analysis confirms $1/L$ as an effective threshold choice, striking a balance between noise suppression and semantic retention, and demonstrating consistent advantages across both datasets.

\newpage
\begin{table*}[!ht]

    \centering
    \resizebox{1.0\textwidth}{!}{
        \begin{tabular}{ccccccc}
            \toprule
            Method & Acc-7 & Acc-5 & Acc-2 & F1 & MAE & Corr\\
            \midrule
            \multicolumn{7}{c}{Random Missing Rate $r=0$} \\
            \midrule
            MISA & 43.05 & 48.30 & 82.78 / 81.24 & 82.83 / 81.23 & 0.771 & 0.777\\
            Self-MM & 42.81 & \textbf{52.38} & \textbf{85.22} / \textbf{83.24} & \textbf{85.19} / \textbf{83.26} & \textbf{0.720} & \textbf{0.790} \\
            MMIM & \textbf{45.92} & 49.85 & 83.43 / 81.97 & 83.43 / 81.94 & 0.744 & 0.778\\
            TFR-Net & 40.82 & 47.91 & 83.64 / 81.68 & 83.57 / 81.61 & 0.805 & 0.760\\
            CENET & 43.20 & 50.39 & 83.08 / 81.49 & 83.06 / 81.48 & 0.748 & 0.785\\
            ALMT & 42.37 & 48.49 & 84.91 / 82.75 & 85.01 / 82.94 & 0.752 & 0.768\\
            BI-Mamba & 39.21 & 43.15 & 82.77 / 81.20 & 82.82 / 81.18 & 0.814 & 0.740 \\
            LNLN & 40.77 & 46.41 & 83.28 / 79.78 & 83.17 / 79.73 & 0.764 & 0.772 \\
            TF-Mamba & 44.31 & 50.58 & 83.69 / 81.63 & 83.71 / 81.58 & 0.762 & 0.774   \\
            
            \midrule
            \multicolumn{7}{c}{Random Missing Rate $r=0.1$} \\
            \midrule
             MISA & 40.28 & 46.21 & 80.18 / 79.01 & 80.21 / 78.97 & 0.847 & 0.721\\
            Self-MM & 40.33 & \textbf{49.03} & {81.40} / \textbf{80.03} & 81.19 / {80.03} & \textbf{0.812} & 0.728 \\
            MMIM & 42.61 & 46.65 & 79.98 / 78.13 & 79.83 / 77.99 & 0.825 & 0.718 \\
            TFR-Net & 38.63 & 45.82 & 79.27 / 77.99 & 78.70 / 77.61 & 0.872 & 0.705 \\
            CENET & 40.13 & 46.60 & 80.08 / 78.38 & 79.91 / 78.20 & 0.837 & 0.719 \\
            ALMT & 39.84 & 45.48 & 80.90 / 78.67 & 81.15 / 79.08 & 0.843 & 0.703 \\
            BI-Mamba & 38.78 & 42.71 & 81.25 / \textbf{80.03} & 81.34 / \textbf{80.06} & 0.873 & 0.699  \\
            LNLN & 39.26 & 44.70 & 80.99 / 78.96 & 80.86 / 78.92 & 0.834 & 0.714 \\
            TF-Mamba & \textbf{42.86} & {48.40} & \textbf{81.86} / \textbf{80.03} & \textbf{81.87} / 79.97 & {0.824} & \textbf{0.732} \\

            \midrule
            \multicolumn{7}{c}{Random Missing Rate $r=0.2$} \\
            \midrule
            MISA & 36.25 & 41.55 & 77.54 / 76.34 & 77.58 / 76.30 & 0.939 & 0.654\\
            Self-MM & 36.64 & 43.98 & 78.15 / 76.48 & 77.76 / 76.51 & 0.901 & 0.660 \\
            MMIM & 39.07 & 42.66 & 76.42 / 74.54 & 76.12 / 74.22 & 0.918 & 0.651 \\
            TFR-Net & 34.70 & 40.13 & 74.70 / 73.52 & 73.57 / 72.70 & 0.987 & 0.622 \\
            CENET & 38.00 & 42.32 & 77.49 / 74.64 & 77.35 / 74.28 & 0.916 & 0.654 \\
            ALMT & 35.33 & 40.33 & 77.64 / 75.70 & 77.94 / 76.24 & 0.927 & 0.645 \\
            BI-Mamba & 33.24 & 39.36 & 78.35 / 77.26 & 78.47 / 77.31 & 0.948 & 0.648  \\
            LNLN & 37.22 & 42.13 & 78.51 / 75.95 & 78.42 / 75.95 & 0.908 & 0.653 \\
            TF-Mamba & \textbf{39.21} & \textbf{44.75} & \textbf{80.49} / \textbf{79.15} & \textbf{80.56} / \textbf{79.17} & \textbf{0.879} & \textbf{0.693} \\

            \midrule
            \multicolumn{7}{c}{Random Missing Rate $r=0.3$} \\
            \midrule
            MISA & 34.60 & 38.97 & 75.76 / 74.54 & 75.82 / 74.51 & 0.989 & 0.618\\
            Self-MM & 34.89 & 40.67 & 76.37 / 74.98 & 75.68 / 74.94 & 0.967 & 0.614\\
            MMIM & {36.83} & 40.43 & 74.08 / 71.91 & 73.47 / 71.28 & 0.974 & 0.612\\
            TFR-Net & 32.55 & 38.34 & 72.36 / 71.28 & 70.12 / 69.58 & 1.065 & 0.572\\
            CENET & 34.74 & 38.97 & 76.83 / 72.01 & 76.56 / 71.30 & 0.983 & 0.605\\
            ALMT & 33.04 & 37.17 & 75.15 / 72.94 & 75.51 / 73.66 & 0.992 & 0.596\\
            BI-Mamba & 33.38 & 37.17 & 75.91 / 75.07 & 76.06 / 75.14 & 0.998 & 0.597 \\
            LNLN & 36.44 & 40.62 & 76.12 / 74.68 & 76.02 / 74.69 & 0.963 & 0.600 \\
            TF-Mamba & \textbf{37.76} & \textbf{42.27} & \textbf{77.74} / \textbf{76.53} & \textbf{77.85} / \textbf{76.56} & \textbf{0.932} & \textbf{0.645} \\

            \midrule
            \multicolumn{7}{c}{Random Missing Rate $r=0.4$} \\
            \midrule
            MISA & 32.65 & 35.37 & 73.88 / 72.59 & 73.88 / 72.49 & 1.041 & 0.585\\
            Self-MM & 31.20 & 36.30 & 73.17 / 71.96 & 71.74 / 71.75 & 1.027 & 0.579\\
            MMIM & 33.38 & 35.76 & 70.84 / 68.90 & 69.69 / 67.80 & 1.034 & 0.576\\
            TFR-Net & 30.17 & 35.76 & 68.75 / 67.74 & 64.71 / 64.41 & 1.142 & 0.537\\
            CENET & 32.26 & 36.15 & 73.38 / 71.53 & 72.75 / 70.26 & 1.031 & 0.574\\
            ALMT & 31.44 & 35.03 & 73.12 / 71.14 & 73.85 / 72.47 & 1.045 & 0.560\\
            BI-Mamba & 32.22 & 35.86 & 74.54 / 73.91 & 74.70 / 73.97 & 1.040 & 0.555 \\
            LNLN & {35.42} & 38.87 & 75.30 / 73.67 & 75.23 / 73.66 & 0.995 & 0.575 \\
            TF-Mamba & \textbf{35.86} & \textbf{40.23} & \textbf{76.07} / \textbf{75.22} & \textbf{76.16} / \textbf{75.25} & \textbf{0.961} & \textbf{0.617} \\
            \bottomrule
        \end{tabular}

        \hfill
        
        \begin{tabular}{ccccccc}
            \toprule
            Method & Acc-7 & Acc-5 & Acc-2 & F1 & MAE & Corr\\
            \midrule
            \multicolumn{7}{c}{Random Missing Rate $r=0.5$} \\
            \midrule
            MISA & 28.14 & 30.61 & 70.53 / 69.34 & 70.50 / 69.20 & 1.124 & 0.519\\
            Self-MM & 26.97 & 31.39 & 67.43 / 67.54 & 64.27 / 66.81 & 1.129 & 0.503\\
            MMIM & 28.23 & 29.89 & 68.09 / 66.52 & 66.15 / 64.59 & 1.128 & 0.501\\
            TFR-Net & 25.85 & 30.71 & 64.83 / 63.02 & 58.04 / 56.64 & 1.270 & 0.443\\
            CENET & 28.33 & 30.90 & 72.46 / 66.08 & 71.10 / 63.50 & 1.130 & 0.496\\
            ALMT & 28.42 & 31.25 & 68.24 / 65.94 & 69.74 / 68.54 & 1.138 & 0.485\\
            BI-Mamba & 31.78 & 34.40 & 73.48 / 73.32 & 73.64 / 73.39 & 1.105 & 0.497 \\
            LNLN & \textbf{33.82} & 37.36 & 73.12 / 72.06 & 72.94 / 72.02 & 1.075 & 0.509 \\
            TF-Mamba & 33.67 & \textbf{37.46} & \textbf{75.00} / \textbf{74.20} & \textbf{75.15} / \textbf{74.27} & \textbf{1.044} & \textbf{0.557} \\

            \midrule
            \multicolumn{7}{c}{Random Missing Rate $r=0.6$} \\
            \midrule
            MISA & 24.68 & 27.12 & 66.97 / 65.84 & 66.94 / 65.69 & 1.200 & 0.441\\
            Self-MM & 24.34 & 27.31 & 63.47 / 63.36 & 58.94 / 62.07 & 1.209 & 0.425\\
            MMIM & 25.41 & 27.11 & 63.67 / 62.49 & 60.87 / 59.48 & 1.208 & 0.418\\
            TFR-Net & 24.05 & 28.33 & 61.64 / 59.47 & 52.44 / 50.53 & 1.371 & 0.363\\
            CENET & 24.54 & 26.53 & 67.58 / 61.47 & 64.87 / 57.86 & 1.215 & 0.415\\
            ALMT & 25.41 & 27.36 & 64.53 / 62.15 & 66.81 / 65.87 & 1.214 & 0.407\\
            BI-Mamba & 29.88 & 29.01 & 67.38 / 67.35 & 67.41 / 66.75 & 1.193 & 0.409 \\
            LNLN & 29.98 & {33.23} & 67.58 / 66.71 & 67.41 / 67.53 & 1.177 & 0.421\\
            TF-Mamba & \textbf{30.76} & \textbf{33.53} & \textbf{68.60} / \textbf{68.37} & \textbf{68.79} / \textbf{68.45} & \textbf{1.127} & \textbf{0.487} \\

            \midrule
            \multicolumn{7}{c}{Random Missing Rate $r=0.7$} \\
            \midrule
            MISA & 21.14 & 23.27 & 65.09 / 63.89 & 65.07 / 63.74 & 1.257 & 0.381\\
            Self-MM & 20.70 & 23.81 & 61.74 / 61.46 & 55.11 / 58.97 & 1.271 & 0.339\\
            MMIM & 22.35 & 24.00 & 61.23 / 59.18 & 57.15 / 54.36 & 1.267 & 0.342\\
            TFR-Net & 23.71 & 26.92 & 59.91 / 57.34 & 48.41 / 45.48 & 1.454 & 0.276\\
            CENET & 22.35 & 23.57 & 63.82 / 59.43 & 53.79 / 54.22 & 1.269 & 0.335\\
            ALMT & 23.71 & 24.97 & 61.84 / 59.67 & 65.30 / 65.19 & 1.266 & 0.336\\
            BI-Mamba & \textbf{27.41} & 27.26 & 64.79 / 65.01 & 64.72 / 64.81 & 1.250 & 0.333 \\
            LNLN & {27.26} & \textbf{30.52} & 64.94 / 63.95 & 64.85 / 63.98 & 1.244 & 0.341\\
            TF-Mamba & {27.26} & 29.30 &\textbf{67.23} / \textbf{66.91} & \textbf{67.41} / \textbf{66.98} & \textbf{1.196} & \textbf{0.411} \\

            \midrule
            \multicolumn{7}{c}{Random Missing Rate $r=0.8$} \\
            \midrule
            MISA & 19.92 & 20.99 & {63.56} / 62.24 & 63.16 / 61.67 & 1.311 & 0.321\\
            Self-MM & 19.29 & 22.11 & 59.55 / 58.26 & 49.98 / 53.56 & 1.313 & 0.282\\
            MMIM & 20.26 & 21.77 & 58.33 / 55.30 & 52.46 / 47.89 & 1.312 & 0.287\\
            TFR-Net & 23.23 & 27.70 & 58.49 / 55.98 & 44.70 / 41.88 & 1.497 & 0.155\\
            CENET & 21.14 & 21.67 & 60.93 / 57.53 & 54.68 / 50.80 & 1.314 & 0.274\\
            ALMT & 23.13 & 23.66 & 60.37 / 58.31 & \textbf{65.45} / \textbf{66.14} & 1.310 & 0.273\\
            BI-Mamba & 24.20 & \textbf{26.38} & 61.13 / 60.93 & 61.10 / 60.78 & 1.289 & 0.300\\
            LNLN & 23.13 & 25.70 & 62.04 / 60.40 & 61.85 / 60.32 & 1.294 & 0.288 \\
            TF-Mamba & \textbf{24.93} & \textbf{26.38} & \textbf{63.57} / \textbf{63.12} & 63.77 / 63.20 & \textbf{1.258} & \textbf{0.353}  \\

            \midrule
            \multicolumn{7}{c}{Random Missing Rate $r=0.9$} \\
            \midrule
            MISA & 17.78 & 18.41 & 58.64 / {58.21} & 56.84 / 56.19 & 1.369 & \textbf{0.226}\\
            Self-MM & 18.32 & 19.78 & 58.59 / 55.25 & 46.16 / 47.46 & 1.353 & 0.197\\
            MMIM & 18.95 & 19.53 & 55.29 / 51.65 & 47.33 / 40.89 & 1.357 & 0.186\\
            TFR-Net & 21.67 & \textbf{25.12} & 57.93 / 55.44 & 43.01 / 40.18 & 1.534 & 0.155\\
            CENET & 19.15 & 19.10 & {58.99} / 54.76 & 50.01 / 46.58 & 1.357 & 0.181\\
            ALMT & {20.31} & 20.50 & 57.32 / 56.66 & \textbf{64.92} / \textbf{67.82} & \textbf{1.349} & 0.205\\
            BI-Mamba & 21.87 & 24.93 & 57.77 / 57.14 & 58.03 / 57.25 & 1.354 & 0.203 \\
            LNLN & 22.01 & 22.93 & 57.17 / 54.91 & 56.38 / 54.13 & 1.371 & 0.164 \\
            TF-Mamba & \textbf{22.89} & 24.49 & \textbf{60.37} / \textbf{60.20} & 60.59 / 60.30 & 1.363 & 0.215  \\
            \bottomrule
        \end{tabular}
        }
        \caption{Details of robust comparison on MOSI with different random missing rates.}
            \label{tab:Details_Robust_Compare_MOSI}
\end{table*}

\begin{table*}[!ht]
    \centering
    \resizebox{1.0\textwidth}{!}{
        \begin{tabular}{ccccccc}
            \toprule
            Method & Acc-7 & Acc-5 & Acc-2 & F1 & MAE & Corr\\
            \midrule
            \multicolumn{7}{c}{Random Missing Rate $r=0$} \\
            \midrule
            MISA & 51.79 & 53.85 & 85.28 / 84.10 & 85.10 / 83.75 & 0.552 & 0.759 \\
            Self-MM & 53.89 & 55.72 & 85.34 / \textbf{84.68} & 85.11 / \textbf{84.66} & \textbf{0.531} & 0.764 \\
            MMIM & 50.76 & 53.04 & 83.53 / 81.65 & 83.39 / 81.41 & 0.576 & 0.724 \\
            TFR-Net & 53.71 & 47.91 & 84.96 / 84.65 & 84.71 / 84.34 & 0.550 & 0.745 \\
            CENET & \textbf{54.39} & \textbf{56.12} & 85.49 / 82.30 & 85.41 / 82.60 & \textbf{0.531} & \textbf{0.770} \\
            ALMT & 52.18 & 53.89 & \textbf{85.62} / 83.99 & \textbf{85.69} / 84.53 & 0.542 & 0.752 \\
            BI-Mamba & 48.40 & 49.71 & 83.65 / 81.84 & 83.60 / 81.77 & 0.620 & 0.677 \\
            LNLN & 50.66 & 51.94 & 84.14 / 83.61 & 84.53 / 84.02 & 0.572 & 0.735 \\
            TF-Mamba & 52.26 & 53.83 & 83.82 / 82.89 & 83.71 / 82.92 & 0.556 & 0.748 \\
            
            \midrule
            \multicolumn{7}{c}{Random Missing Rate $r=0.1$} \\
            \midrule
            MISA & 50.13 & 51.34 & 82.21 / 82.28 & 81.28 / 80.79 & 0.598 & 0.722 \\
            Self-MM & 51.80 & 53.18 & 83.03 / \textbf{83.79} & 82.43 / \textbf{83.23} & 0.564 & 0.725 \\
            MMIM & 49.09 & 51.19 & 82.00 / 81.09 & 81.57 / 80.15 & 0.602 & 0.696 \\
            TFR-Net & 52.29 & 45.82 & 82.92 / 83.31 & 82.25 / 82.40 & 0.573 & 0.715 \\
            CENET & \textbf{52.83} & \textbf{54.23} & 83.75 / 82.41 & 83.42 / 82.34 & \textbf{0.556} & \textbf{0.739} \\
            ALMT & 49.98 & 51.38 & \textbf{84.14} / 82.84 & \textbf{84.23} / {83.04} & 0.583 & 0.718 \\
            BI-Mamba & 48.36 & 49.43 & 82.97 / 80.79 & 82.92 / 80.61 & 0.630 & 0.662 \\
            LNLN & 49.96 & 51.25 & 83.32 / 82.73 & 83.66 / 82.91 & 0.591 & 0.712 \\
            TF-Mamba & 50.53 & 52.07 & 83.16 / 82.68 & 83.03 / 82.69 & 0.570 & 0.730\\

            \midrule
            \multicolumn{7}{c}{Random Missing Rate $r=0.2$} \\
            \midrule
            MISA  & 47.24 & 47.66 & 77.84 / 79.93 & 75.56 / 76.88 & 0.659 & 0.674 \\
            Self-MM & 49.44 & 50.51 & 80.84 / \textbf{82.33} & 79.76 / 81.17 & 0.604 & 0.678 \\
            MMIM & 46.27 & 47.99 & 79.93 / 79.66 & 79.08 / 77.68 & 0.642 & 0.653 \\
            TFR-Net & \textbf{51.04} & 40.13 & 80.47 / 81.61 & 79.29 / 79.99 & 0.604 & 0.672 \\
            CENET & {50.72} & \textbf{51.85} & 81.46 / 81.62 & 80.78 / 81.17 & 0.590 & 0.698 \\
            ALMT & 46.61 & 47.82 & \textbf{82.71} / 81.65 & \textbf{82.82} / 81.83 & 0.607 & 0.669 \\
            BI-Mamba & 47.91 & 49.26 & 81.56 / 80.30 & 81.51 / 80.03 & 0.644 & 0.641\\
            LNLN & 48.75 & 49.95 & 81.70 / 81.68 & 81.95 / \textbf{81.89} & 0.616 & 0.677 \\
            TF-Mamba & 49.17 & 50.50 & 82.55 / 81.84 & 82.40 / 81.83 & \textbf{0.588} & \textbf{0.710} \\

            \midrule
            \multicolumn{7}{c}{Random Missing Rate $r=0.3$} \\
            \midrule
            MISA  & 43.99 & 43.40 & 73.32 / 77.28 & 68.91 / 72.25 & 0.724 & 0.615 \\
            Self-MM & 47.23 & 48.07 & 77.63 / 79.99 & 75.69 / 77.74 & 0.653 & 0.610 \\
            MMIM & 43.25 & 44.73 & 77.08 / 77.79 & 75.46 / 74.49 & 0.690 & 0.597 \\
            TFR-Net  & \textbf{48.75} & 38.34 & 77.48 / 79.29 & 75.43 / 76.52 & 0.650 & 0.604 \\
            CENET & 48.49 & \textbf{49.37} & 78.65 / 80.02 & 77.34 / 78.94 & 0.636 & 0.640 \\
            ALMT  & 43.04 & 44.05 & \textbf{80.94} / 79.94 & \textbf{81.15} / 80.20 & 0.632 & 0.598 \\
            BI-Mamba & 46.98 & 48.38 & 80.27 / 79.48 & 80.17 / 79.13 & 0.662 & 0.615 \\
            LNLN  & 47.36 & 48.40 & 80.11 / 80.45 & 80.44 / 80.91 & 0.648 & 0.629 \\
            TF-Mamba & 47.89 & 49.00 & 80.82 / \textbf{81.22} & 80.58 / \textbf{81.10} & \textbf{0.613 }& \textbf{0.675} \\

            \midrule
            \multicolumn{7}{c}{Random Missing Rate $r=0.4$} \\
            \midrule
            MISA & 40.87 & 39.53 & 70.46 / 75.04 & 64.02 / 67.93 & 0.780 & 0.561 \\
            Self-MM & 44.40 & 45.04 & 75.02 / 78.09 & 72.01 / 74.48 & 0.694 & 0.554 \\
            MMIM & 40.84 & 41.86 & 74.56 / 76.15 & 71.98 / 71.40 & 0.732 & 0.542 \\
            TFR-Net & 46.70 & 35.76 & 74.74 / 77.65 & 71.67 / 73.71 & 0.688 & 0.548 \\
            CENET & \textbf{47.12} & 47.74 & 76.03 / 78.57 & 73.87 / 76.75 & 0.678 & 0.587 \\
            ALMT & 40.40 & 41.21 & 79.40 / 79.16 & 79.68 / 79.50 & 0.651 & 0.536 \\
            BI-Mamba & 45.98 & 46.53 & 78.62 / 78.58 & 78.55 / 78.26 & 0.678 & 0.588 \\
            LNLN & 45.99 & 46.88 & 78.49 / 79.70 & 78.98 / \textbf{80.46} & 0.673 & 0.592 \\
            TF-Mamba & 46.73 & \textbf{47.80} & \textbf{80.02} / \textbf{80.02} & \textbf{79.80} / 79.80 & \textbf{0.639} & \textbf{0.638} \\
            \bottomrule
        \end{tabular}

        \hfill
        
        \begin{tabular}{ccccccc}
            \toprule
            Method & Acc-7 & Acc-5 & Acc-2 & F1 & MAE & Corr\\
            \midrule
            \multicolumn{7}{c}{Random Missing Rate $r=0.5$} \\
            \midrule
            MISA & 38.12 & 36.05 & 67.38 / 73.21 & 58.38 / 64.14 & 0.834 & 0.492 \\
            Self-MM & 42.70 & 43.14 & 71.97 / 75.81 & 67.40 / 70.38 & 0.733 & 0.477 \\
            MMIM & 38.68 & 39.21 & 71.75 / 74.45 & 67.70 / 67.96 & 0.775 & 0.470 \\
            TFR-Net & 45.00 & 30.71 & 71.53 / 75.69 & 66.88 / 70.07 & 0.730 & 0.471 \\
            CENET  & {45.12} & 45.52 & 73.33 / 77.16 & 69.80 / 74.14 & 0.720 & 0.515 \\
            ALMT & 37.82 & 38.34 & {77.40} / 77.48 & {77.73} / 77.80 & {0.683} & 0.461 \\
            BI-Mamba & 45.44 & 46.21 & 77.41 / 77.48 & 77.25 / 77.17 & 0.699 & 0.550 \\
            LNLN & 44.90 & 45.59 & 76.44 / \textbf{78.10} & 77.23 / \textbf{79.30} & 0.710 & {0.529} \\
            TF-Mamba & \textbf{45.68} & \textbf{46.60} & \textbf{78.59} / 77.94 & \textbf{78.34} / 77.78 & \textbf{0.676} & \textbf{0.583} \\

            \midrule
            \multicolumn{7}{c}{Random Missing Rate $r=0.6$} \\
            \midrule
            MISA & 36.16 & 33.30 & 65.55 / 72.30 & 54.64 / 62.12 & 0.875 & 0.415 \\
            Self-MM & 41.47 & 41.75 & 69.33 / 73.93 & 63.01 / 66.76 & 0.762 & 0.401 \\
            MMIM & 37.13 & 37.48 & 68.83 / 73.16 & 63.09 / 65.43 & 0.808 & 0.402 \\
            TFR-Net & 43.88 & 28.33 & 68.80 / 74.05 & 62.51 / 67.07 & 0.762 & 0.397 \\
            CENET & \textbf{44.45} & {44.64} & 70.50 / 75.39 & 65.27 / 70.86 & 0.749 & 0.446 \\
            ALMT & 35.99 & 36.30 & {74.98} / 76.26 & {75.44} / 76.71 & {0.710} & 0.395 \\
            BI-Mamba & 43.16 & 43.92 & 75.21 / 75.17 & 74.90 / 74.94 & 0.730 & 0.512 \\
            LNLN  & 43.52 & 44.00 & 73.82 / \textbf{76.50} & 75.03 / \textbf{78.33} & 0.736 & {0.471} \\
            TF-Mamba & 43.96 & \textbf{44.73} & \textbf{75.89} / 75.77 & \textbf{75.63} / 75.59 & \textbf{0.709} & \textbf{0.534} \\

            \midrule
            \multicolumn{7}{c}{Random Missing Rate $r=0.7$} \\
            \midrule
            MISA & 34.54 & 31.21 & 64.28 / 71.71 & 51.82 / 60.65 & 0.906 & 0.344 \\
            Self-MM & 39.93 & 40.12 & 66.79 / 72.55 & 58.05 / 63.45 & 0.786 & 0.329 \\
            MMIM & 35.25 & 35.47 & 66.89 / 72.26 & 58.90 / 63.26 & 0.834 & 0.341 \\
            TFR-Net & 42.91 & 26.92 & 66.64 / 72.77 & 58.32 / 64.02 & 0.786 & 0.322 \\
            CENET & \textbf{43.93} & \textbf{44.03} & 67.50 / 73.39 & 59.88 / 67.02 & 0.776 & 0.384 \\
            ALMT & 34.78 & 34.95 & {71.62} / 73.98 & 72.24 / 74.54 & \textbf{0.743} & 0.315 \\
            BI-Mamba & 42.20 & 43.16 & 72.04 / 73.45 & 71.44 / 73.04 & 0.754 & 0.466 \\
            LNLN & 42.22 & 42.56 & 71.55 / \textbf{74.74} & \textbf{73.49} / \textbf{77.40} & 0.762 & {0.408} \\
            TF-Mamba & 42.78 & 43.40 & \textbf{73.50} / 73.49 & 73.26 / 73.19 & 0.747 & \textbf{0.469} \\

            \midrule
            \multicolumn{7}{c}{Random Missing Rate $r=0.8$} \\
            \midrule
            MISA & 33.29 & 29.51 & 63.43 / 71.30 & 49.95 / 59.69 & 0.927 & 0.267 \\
            Self-MM & 38.69 & 38.78 & 65.07 / 71.83 & 54.44 / 61.49 & 0.805 & 0.259 \\
            MMIM & 33.64 & 33.71 & 64.97 / 71.57 & 54.76 / 61.45 & 0.858 & 0.269 \\
            TFR-Net & 42.23 & 27.70 & 65.05 / 71.95 & 54.91 / 61.82 & 0.807 & 0.241 \\
            CENET & \textbf{42.71} & \textbf{42.74} & 65.88 / 72.16 & 56.80 / 64.67 & 0.798 & 0.316 \\
            ALMT  & 34.01 & 34.09 & 68.15 / 71.48 & 69.12 / 72.28 & \textbf{0.774} & 0.231 \\
            BI-Mamba & 42.05 & 41.40 & 70.12 / 71.35 & 69.08 / 70.90 & 0.775 & 0.421 \\
            LNLN  & 40.76 & 40.97 & {68.62} / \textbf{72.86} & \textbf{71.83} / \textbf{76.80} & 0.791 & {0.325} \\
            TF-Mamba & 40.37 & 40.93 & \textbf{70.34} / 71.60 & 70.16 / 71.27 & 0.786 & \textbf{0.408}\\

            \midrule
            \multicolumn{7}{c}{Random Missing Rate $r=0.9$} \\
            \midrule
            MISA & 32.29 & 28.03 & 62.95 / 71.07 & 48.80 / 59.12 & 0.941 & 0.180 \\
            Self-MM & 37.46 & 37.50 & 63.85 / 71.24 & 51.32 / 59.72 & 0.821 & 0.188 \\
            MMIM & 32.61 & 32.67 & 63.69 / 71.10 & 51.26 / 59.99 & 0.877 & 0.197 \\
            TFR-Net & 41.73 & 25.12 & 63.64 / 71.34 & 52.02 / 59.99 & 0.820 & 0.175 \\
            CENET & \textbf{42.08} & \textbf{42.08} & 64.14 / 70.42 & 54.27 / 62.33 & 0.814 & {0.254} \\
            ALMT & 34.40 & 34.40 & 61.41 / 68.65 & 63.32 / 69.83 & \textbf{0.810} & 0.138 \\
            BI-Mamba & 40.74 & 39.56 & 66.32 / 68.77 & 64.11 / 67.96 & 0.817 & \textbf{0.321} \\
            LNLN & 40.10 & 40.19 & \textbf{64.83} / \textbf{71.51} & \textbf{70.60} / \textbf{77.52} & 0.820 & 0.221 \\
            TF-Mamba & 37.24 & 37.56 & 64.75 / 68.68 & 64.87 / 68.18 & 0.851 & {0.291} \\
            \bottomrule
        \end{tabular}
        }
        \caption{Details of robust comparison on MOSEI with different random missing rates.}
            \label{tab:Details_Robust_Compare_MOSEI}

\end{table*}

\begin{table*}[!ht]
    \centering
    \small
    \resizebox{1.0\textwidth}{!}{
        \begin{tabular}{ccccccc}
            \toprule
            Method & Acc-5 & Acc-3 & Acc-2 & F1 & MAE & Corr\\
            \midrule
            \multicolumn{7}{c}{Random Missing Rate $r=0$} \\
            \midrule
            MISA & 40.55 & 63.38 & 78.19 & 77.22 & 0.449 & 0.576 \\
            Self-MM & 40.77 & \textbf{64.92} & 78.26 & 78.00 & \textbf{0.421} & \textbf{0.584} \\
            MMIM & 37.42 & 60.69 & 75.42 & 73.10 & 0.475 & 0.528 \\
            TFR-Net & 33.85 & 54.12 & 69.15 & 58.44 & 0.562 & 0.254 \\
            CENET & 23.85 & 54.05 & 68.71 & 57.82 & 0.578 & 0.137 \\
            ALMT & 23.41 & 54.78 & 75.64 & 76.27 & 0.527 & 0.536 \\
            BI-Mamba & \textbf{41.58} & 63.02 & 76.59 & 76.93 & 0.433 & 0.574\\
            LNLN & 38.51 & 61.78 & 77.68 & 77.22 & 0.448 & 0.561 \\
            TF-Mamba & 37.86 & 61.93 & \textbf{79.65} & \textbf{78.92} & 0.441 & 0.548 \\

            \midrule
            \multicolumn{7}{c}{Random Missing Rate $r=0.1$} \\
            \midrule
            MISA & 38.88 & 63.02 & 77.39 & 75.82 & 0.461 & 0.561 \\
            Self-MM & 40.26 & \textbf{63.53} & 77.32 & 76.76 & \textbf{0.433} & \textbf{0.563} \\
            MMIM & 37.27 & 60.90 & 74.25 & 72.08 & 0.473 & 0.529 \\
            TFR-Net & 30.12 & 53.25 & 68.85 & 59.38 & 0.596 & 0.203 \\     
            CENET & 22.83 & 53.98 & 68.57 & 57.36 & 0.580 & 0.136 \\
            ALMT & 22.10 & 55.14 & 74.40 & 75.19 & 0.530 & 0.537 \\
            BI-Mamba & \textbf{40.70} & 62.80 & 74.84 & 75.21 & 0.448 & 0.561 \\
            LNLN & 37.27 & 62.80 & 78.12 & 77.54 & 0.450 & 0.554 \\
            TF-Mamba & 36.98 & 62.36 & \textbf{79.65} & \textbf{78.79} & 0.445 & 0.550 \\
            
            \midrule
            \multicolumn{7}{c}{Random Missing Rate $r=0.2$} \\
            \midrule
            MISA & 38.15 & 59.23 & 74.33 & 71.70 & 0.489 & 0.490 \\
            Self-MM & \textbf{38.37} & \textbf{61.71} & 74.98 & 73.71 & 0.464 & 0.500 \\
            MMIM & 37.27 & 57.33 & 72.36 & 69.80 & 0.504 & 0.460 \\
            TFR-Net & 29.03 & 53.61 & 68.64 & 59.74 & 0.619 & 0.191 \\
            CENET & 22.25 & 54.20 & 68.57 & 57.64 & 0.583 & 0.132 \\
            ALMT & 21.08 & 53.17 & 72.65 & 73.90 & 0.541 & 0.485 \\
            BI-Mamba & 37.20 & 59.74 & 72.65 & 72.56 & 0.472 & 0.506 \\
            LNLN & 35.59 & 60.69 & 76.29 & 75.53 & 0.468 & \textbf{0.509 }\\
            TF-Mamba & 38.29 & 61.05 & \textbf{78.77} & \textbf{77.88} & \textbf{0.459} & 0.507\\

            \midrule
            \multicolumn{7}{c}{Random Missing Rate $r=0.3$} \\
            \midrule
            MISA & 36.40 & 59.30 & 74.11 & 70.40 & 0.505 & 0.464 \\
            Self-MM & 37.93 & \textbf{59.81} & 74.76 & 72.85 & 0.474 & 0.487 \\
            MMIM & 37.71 & 58.06 & 72.36 & 69.52 & 0.512 & 0.436 \\
            TFR-Net & 27.64 & 52.30 & 68.42 & 59.88 & 0.640 & 0.182 \\
            CENET & 21.44 & 54.05 & 68.42 & 57.41 & 0.578 & 0.175 \\
            ALMT & 20.35 & 50.62 & 72.06 & 73.64 & 0.546 & 0.469 \\
            BI-Mamba & 33.04 & 58.21 & 72.21 & 72.39 & 0.505 & 0.451 \\
            LNLN & 36.32 & 58.94 & 75.35 & 73.60 & 0.475 & \textbf{0.502} \\
            TF-Mamba & \textbf{39.82} & 58.42 & \textbf{75.93} & \textbf{74.72} & \textbf{0.468} & 0.485\\

            \midrule
            \multicolumn{7}{c}{Random Missing Rate $r=0.4$} \\
            \midrule
            MISA & 34.86 & 57.33 & 72.87 & 67.52 & 0.523 & 0.436 \\
            Self-MM & 34.57 & 58.28 & 73.30 & 70.36 & 0.482 & \textbf{0.479} \\
            MMIM & 34.57 & 55.36 & 69.95 & 66.49 & 0.533 & 0.399 \\
            TFR-Net & 25.31 & 51.86 & 67.91 & 59.16 & 0.664 & 0.176 \\
            CENET & 22.54 & 54.12 & 68.49 & 57.68 & 0.583 & 0.141 \\
            ALMT & 19.91 & 49.45 & 70.75 & 72.97 & 0.549 & 0.470 \\
            BI-Mamba & 32.17 & 55.58 & 70.90 & 70.75 & 0.509 & 0.419\\
            LNLN & 35.74 & 58.86 & 74.18 & 71.38 & 0.488 & 0.478 \\
            TF-Mamba & \textbf{40.04} & \textbf{60.39} & \textbf{75.49} & \textbf{73.72} & \textbf{0.470} & 0.477 \\
            \bottomrule
        \end{tabular}

        \hfill
        
        \begin{tabular}{ccccccc}
            \toprule
            Method & Acc-5 & Acc-3 & Acc-2 & F1 & MAE & Corr\\
            \midrule
            \multicolumn{7}{c}{Random Missing Rate $r=0.5$} \\
            \midrule
            MISA & 30.56 & 54.78 & 71.26 & 64.16 & 0.552 & 0.367 \\
            Self-MM & 32.02 & 53.90 & 71.41 & 67.11 & 0.517 & 0.390 \\
            MMIM & 33.41 & 52.37 & 68.49 & 64.81 & 0.553 & 0.336 \\
            TFR-Net & 24.65 & 52.37 & 67.47 & 58.66 & 0.685 & 0.171 \\
            CENET & 23.12 & 54.05 & 68.71 & 57.92 & 0.588 & 0.107 \\
            ALMT & 18.38 & 47.12 & 68.27 & 71.22 & 0.563 & 0.395 \\
            BI-Mamba & 30.42 & 56.46 & 71.33 & 71.06 & 0.540 & 0.334 \\
            LNLN & 35.81 & 57.70 & 73.74 & 69.84 & 0.515 & 0.416 \\
            TF-Mamba & \textbf{37.20} & \textbf{58.64} & \textbf{75.71} & \textbf{73.61} & \textbf{0.495} & \textbf{0.424} \\
            
            \midrule
            \multicolumn{7}{c}{Random Missing Rate $r=0.6$} \\
            \midrule
            MISA & 27.72 & 53.97 & 70.46 & 61.81 & 0.578 & 0.286 \\
            Self-MM & 29.10 & 51.86 & 70.02 & 64.21 & 0.548 & 0.313 \\
            MMIM & 29.18 & 49.31 & 67.91 & 63.86 & 0.578 & 0.270 \\
            TFR-Net & 24.80 & 52.59 & 67.03 & 58.30 & 0.696 & 0.157 \\
            CENET & 22.46 & 53.69 & 69.00 & 58.64 & 0.592 & 0.102 \\
            ALMT & 18.67 & 43.69 & 66.81 & \textbf{70.69} & 0.574 & 0.322 \\
            BI-Mamba & 26.04 & 50.98 & 68.49 & 67.84 & 0.554 & 0.281 \\
            LNLN & 32.31 & 54.85 & 71.77 & 65.57 & 0.543 & 0.345 \\
            TF-Mamba & \textbf{32.82} & \textbf{54.92} & \textbf{73.09} & 69.44 & \textbf{0.523} & \textbf{0.370 }\\
            
            \midrule
            \multicolumn{7}{c}{Random Missing Rate $r=0.7$} \\
            \midrule
            MISA & 24.87 & 52.52 & {69.95} & 59.54 & 0.601 & 0.167 \\
            Self-MM & 25.53 & 50.62 & 69.58 & 62.28 & 0.571 & 0.198 \\
            MMIM & 28.59 & 46.53 & 66.89 & 62.23 & 0.595 & 0.190 \\
            TFR-Net & 23.78 & 52.30 & 67.18 & 58.15 & 0.707 & 0.163 \\
            CENET & 21.81 & \textbf{53.32} & 67.69 & 57.87 & 0.599 & 0.070 \\
            ALMT & 18.02 & 38.66 & 65.57 & \textbf{70.27} & 0.586 & 0.218 \\
            BI-Mamba & 24.95 & 45.08 & 66.30 & 63.60 & 0.605 & 0.148 \\
            LNLN & \textbf{29.91} & 50.47 & 70.17 & 61.28 & \textbf{0.566} & 0.236 \\
            TF-Mamba & 28.45 & 48.80 & \textbf{71.77} & 67.46 & 0.570 & \textbf{0.248} \\

            \midrule
            \multicolumn{7}{c}{Random Missing Rate $r=0.8$} \\
            \midrule
            MISA & 22.69 & 52.22 & 69.37 & 57.82 & 0.610 & 0.092 \\
            Self-MM & 22.03 & 50.77 & 69.51 & 60.68 & \textbf{0.585} & 0.138 \\
            MMIM & 22.32 & 44.35 & 65.28 & 60.53 & 0.607 & 0.145 \\
            TFR-Net & 22.97 & \textbf{52.74} & 67.54 & 57.55 & 0.721 & 0.100 \\
            CENET & 21.73 & 52.15 & 67.47 & 58.44 & 0.599 & 0.074 \\
            ALMT & 18.60 & 34.06 & 64.19 & \textbf{69.64} & 0.597 & 0.133 \\
            BI-Mamba & 26.04 & 50.11 & 66.96 & 61.71 & 0.607 & 0.129\\
            LNLN & 25.82 & 49.16 & 69.66 & 59.42 & 0.587 & \textbf{0.164}  \\
            TF-Mamba & \textbf{26.48} & 46.39 & \textbf{71.55} & 66.58 & 0.607 & 0.139\\

            \midrule
            \multicolumn{7}{c}{Random Missing Rate $r=0.9$} \\
            \midrule
            MISA & 20.64 & 52.95 & {69.22} & 57.01 & 0.617 & 0.041 \\
            Self-MM & 22.17 & 52.15 & 68.92 & 58.32 & \textbf{0.586} & 0.111 \\
            MMIM & 20.35 & 42.67 & 65.72 & 59.64 & 0.610 & 0.096 \\
            TFR-Net & 23.05 & \textbf{53.76} & 69.08 & 57.71 & 0.721 & 0.088 \\
            CENET & 20.86 & 48.07 & 65.72 & 58.18 & 0.609 & -0.002 \\
            ALMT & 19.47 & 26.91 & 66.23 & \textbf{73.76} & 0.596 & 0.076 \\
            BI-Mamba & \textbf{26.91} & {47.48} & 67.61 & 60.58 & 0.627 & 0.049  \\
            LNLN & 23.49 & 44.86 & \textbf{69.29} & 57.05 & 0.595 & \textbf{0.128} \\
            TF-Mamba & 26.70 & 42.23 & 65.21 & 60.87 & 0.648 & 0.114\\
            \bottomrule
        \end{tabular}
        }
        \caption{Details of robust comparison on SIMS with different random missing rates.}
            \label{tab:Details_Robust_Compare_SIMS}

\end{table*}

\end{document}